\documentclass[pdftex,twocolumn,epjc3]{svjour3}          

\RequirePackage[T1]{fontenc}

\smartqed  

\RequirePackage{graphicx}
\RequirePackage{mathptmx}      
\RequirePackage{flushend}
\RequirePackage[numbers,sort&compress]{natbib}
\RequirePackage[colorlinks,citecolor=blue,urlcolor=blue,linkcolor=blue]{hyperref}
\RequirePackage{amssymb}

\journalname{Eur. Phys. J. C}

\begin{document}
\newcommand{\ti}[1]{\mbox{\tiny{#1}}}
\def\be{\begin{equation}}
\def\ee{\end{equation}}
\def\bea{\begin{eqnarray}}
\def\eea{\end{eqnarray}}
\newcommand{\bt}[1]{\mathbf{\mathtt{#1}}}
\newcommand{\tb}[1]{\textbf{\texttt{#1}}}
\newcommand{\btb}[1]{\textcolor[rgb]{0.00,0.00,1.00}{\tb{#1}}}
\newcommand{\il}{~}
\newcommand{\rtb}[1]{\textcolor[rgb]{1.00,0.00,0.00}{\tb{#1}}}
\newcommand{\ptb}[1]{\textcolor[rgb]{0.50,0.00,0.50}{\tb{#1}}}
\newcommand{\gtb}[1]{\textcolor[rgb]{0.00,0.80,0.50}{\tb{#1}}}
\newcommand{\otb}[1]{\textcolor[rgb]{1.00,0.50,0.25}{\textbf{#1}}}

\title{The ergoregion in the  Kerr spacetime: properties of the equatorial circular motion
}


\author{D. Pugliese\thanksref{e1,addr1,addr2}
        \and
        H. Quevedo\thanksref{e2,addr3,addr4} 
}

\thankstext{e1}{e-mail: d.pugliese@qmul.ac.uk}
\thankstext{e2}{e-mail: quevedo@nucleares.unam.mx}

\institute{Institute of Physics, Faculty of Philosophy \& Science,
  Silesian University in Opava,
 Bezru\v{c}ovo n\'{a}m\v{e}st\'{i} 13, CZ-74601 Opava, Czech Republic\label{addr1}
          \and
          School of Mathematical Sciences, Queen Mary University of London,
Mile End Road, London E1 4NS, United Kingdom \label{addr2}
          \and
          Instituto de Ciencias Nucleares, Universidad Nacional Aut\'onoma de M\'exico,
AP 70543, M\'exico, DF 04510, Mexico\label{addr3}
\and Dipartimento di Fisica, Universit\`a di Roma ``La Sapienza'', Piazzale Aldo Moro 5, I-00185 Roma, Italy\label{addr4}
}

\date{Received: date / Accepted: date}

\maketitle

\begin{abstract}
We investigate in detail the circular motion of test particles on the
equatorial plane of the ergoregion in the Kerr spacetime. We  consider all the
regions  where circular motion is allowed, and we
analyze the stability properties and the energy and angular momentum of
the test particles. We show that the structure of the stability regions
has definite features that make it possible to distinguish between black
holes and naked singularities. The  naked singularity case  presents a very structured non-connected set
of regions of orbital stability, where the presence  of counterrotating particles and  zero angular momentum
particles for a specific class of  naked
singularities is interpreted as due to the presence of a repulsive field
generated by the central source of gravity. In particular, we analyze the effects of the dynamical
structure of the ergoregion  (the union of the orbital regions for different attractor spins) on the behavior of accretion disks around the central source. The properties of the circular motion turn
out to be so distinctive that they allow   the introduction of a complete
classification of Kerr spacetimes, each class of which is characterized by
different physical effects that could be of especial relevance in
observational Astrophysics. {We also    identify    some  special black hole spacetimes  where these effects could be  relevant}.
\end{abstract}

\section{Introduction}
Black holes are very probably the central engines of Quasars, Active Galactic Nuclei,
and  Gamma Ray Bursts. Consequently, the mechanism by which energy is extracted
from them is of great astrophysical interest. While the exact form of this mechanism is
not known, it seems that the effects occurring inside the ergoregion of black holes are essential
for understanding the central engine mechanism \cite{Meier,Fro-Z}.

On the other hand,
since the discussion first presented in \cite{Sha-teu91}, the issue of  creation and  stability of naked singularities has been intensively debated  with different  controversial results \cite{ApTho,J-S09,Jacobson:2010iu,Esitenza}.
{For instance}, the paper \cite{Giacomazzo:2011cv}  addresses the issue of the possible formation of a naked singularity, analyzing the   stability of the progenitor models and investigating the  gravitational collapse
 of differentially rotating neutron stars (in full general relativity). These results do not exclude the possibility that a naked singularity can be
produced as the result of a gravitational collapse, { under quite general conditions  on the progenitor and considering instability processes}.
{
Since practically all of these results are based upon a numerical integration of the corresponding field equations, their interpretation requires a careful analysis. Indeed, the formation of black holes (with singularities inside the horizon)  is usually associated with the existence of trapped surfaces. So, the numerical detection of a singularity without trapped surfaces is usually considered as a proof that the singularity is naked. This, however, is not always true. As has been shown in \cite{Wald:1991zz}, during the formation of a spherically symmetric black hole it is possible to choose a very particular slicing of spacetime such that no trapped surfaces exist. This shows that the non existence of trapped surfaces in the gravitational collapse cannot be considered in general as a proof of the existence of a naked singularity. A more detailed analysis (further numerical integration) is necessary in order to establish that a naked singularity can be formed as the end result of a gravitational collapse. See, for instance, \cite{Joshi-Book} for particular examples. In any case,
it is
therefore interesting to investigate the physical effects associated with the gravitational field of naked singularities.}

In Astrophysics, it is particularly interesting to study the general features of the motion of test particles moving along circular orbits around the central source. In fact, one can imagine a  thin disk made only of test particles  as a hypothetical accretion disk of matter surrounding the central source. Although this is a very idealized model for an accretion disk, one can nevertheless extract some valuable information about the dynamics of particles in the corresponding gravitational field and the amount of energy that can be released by matter when falling into the central mass distribution (see \cite{chandra42} and \cite{Stuchlik:2011zza,
Li:2012ra,
Joshi:2013dva,
Joshi:2012mk,HightPatil,
HightPatilCQG}),  with the equatorial circular geodesics
  being then relevant for the Keplerian accretion disks. Moreover, there is an increasing interest in the properties of the matter dynamics  in the ergoregion,  especially as a possible source of data discriminating between black holes  and super-spinning  objects (see \cite{Pu:Kerr} and \cite{Gariel:2014ara,Kovacs:2010xm,Stuchlik:2013yca}, also
\cite{Pelavas:2000za,Herdeiro:2014jaa}).
In \cite{Pu:Neutral,Pu:Kerr,Pu:Charged,Pu:Class,Pu:KN}, we studied  the  geometric structure of such an idealized accretion disk and how it can depend on the values of the physical parameters that determine the gravitational field. This is also an important issue since it could lead to physical effects that depend on the structure of the accretion disk, with the corresponding possible observational consequences. Indeed, in a series of previous studies \cite{Pu:Neutral,Pu:Kerr,Pu:Charged,Pu:Class,Pu:KN},
it was established that the motion of test particles on the equatorial plane of black hole spacetimes can be used to derive information about the structure of the central source of gravitation (see also discussions in \cite{Schee:2013bya,
Kovacs:2010xm,
Pat2010}
and
\cite{Stute:2003aj}).
In \cite{Pu:Kerr}  and \cite{Pu:KN}, some of the characteristics of the circular motion  in the Kerr and Kerr-Newman spacetimes for  black holes and naked singularities were discussed,  and typical effects of repulsive gravity  were  enlightened in the  naked singularity ergoregion (see also \cite{Gariel:2014ara,
Kovacs:2010xm,
Stuchlik:2013yca} and \cite{Pelavas:2000za,Herdeiro:2014jaa}). Moreover,   it was pointed out that there exists a  dramatic  difference in black holes and naked singularities  with respect to the zero and  negative energy states in circular orbits.
In this work, we clarify and deepen those results, providing a classification of attractor sources on the basis of the properties of this type of dynamics;
 we focus {here} exclusively on the dynamics within the ergoregion, considering bounded and unbounded unstable orbits,
  and investigate the main possible  astrophysical implications with particular attention to the physics of accretion disks.   {Our aim is  to explore the   dynamics  within this region from the point of view of the orbiting matter  to enlighten also the  instabilities  that could give rise to detectable phenomena in Astrophysics.} This analysis  has two essential features. First, from  the methodological standpoint we  introduce the notion of dynamical structure of the ergoregion which is essentially a spacetime region decomposition  based on the  geometric properties investigated here and, secondly,  we connect in a quite natural manner our study to the analysis of configurations of extended toroidal matter in accretion such as accretion disk thick models, where the  equilibrium  and the  unstable states of the system  are mainly governed  by the curvature effects of the geometry.
	The relevance of these studies in relation to the (hydrodynamic) relativistic disk models had been anticipated in \cite{Pugliese:2014bua} where, however, the regions near the static limit remained unexplored. Here, we take back these considerations, analyze the static limit, and focus our attention on equilibrium configurations as well as on the location and  dynamics  of   unstable configurations  with the associated accretion or launch of  Jets.

From the methodological point of view our approach consists in analyzing the behavior of the energy and angular momentum of the test particles as functions of the radial distance and of the intrinsic angular momentum of the central source, which for the  sake of brevity will be often referred as the ``spin" of the source. This procedure allows us to carry out a methodical and detailed physical analysis of all the regions inside the ergoregion where circular motion is allowed. We investigate the dynamics  inside the ergoregion,  defining and characterizing the  regions where  there are  stable orbits, bounded or unbounded  unstable orbits  and, eventually, the regions where particles with negative energies are allowed.
The disjoint union of these  orbital sections  fully covers the ergoregion and the set of regions   defines  the    {\emph{dynamical structure}} or decomposition of the ergoregion,   fully characterizing indeed  the dynamical properties of the orbiting matter.  {We study the dynamical structure of the ergoregion and the behavior of matter in accretion, identifying different scenarios where there may be  potentially detectable effects.
} We discuss the  possible astrophysical implications of this analysis  in terms of the dynamical structure,    particularly, in relation with  the source evolution  and the configuration of the extended matter.
The dynamical structure  plays in fact a specific  and fundamental role in the modelization  of   accretion disks.
The properties of matter in circular configurations, i.e.,  toroidal configurations  of extended matter in accretion,  are in general governed by several factors such as  the pressure,   the  magnetic fields,  the dissipative effects, etc., but the properties of the spacetime structure highlighted by the geodesics motion  are the   basis of any disk model.
In   particular, we refer  here to  the relativistic models of pressure supported disks  where the role of the hydrostatic pressure and the  background are crucial for determining  both the equilibrium phases  and   the gravitational and  hydrostatic  instabilities,  leading to the accretion  or  the launch of Jets  in   funnels of matter  from the instability regions \cite{Abramowicz:2011xu,Lei:2008ui,Pugliese:2014bua}.

In general, one might look at the entire evolution of a thick accretion disk,  from the formation of a thin ring, its growing  to the unstable phases of the accretion, resulting  finally in the formation of Jets of matter from the cusp point, as  the orbiting matter transition in the dynamical structure of the ergoregion. As the disk equatorial plane of symmetry  often coincides with  the symmetric plane of the accreitor, by using the symmetry of the system  we can  reduce  the  analysis to a one-dimensional problem along the radial direction,  the dynamics of the orbiting matter   being then regulated entirely by a couple of  model parameters  so that the main features of the disk dynamics, stability and even the  accretion,  are mainly regulated by the configuration properties on the plane of symmetry. We will see that
from this analysis different classes  of attractors emerge.
Moreover, we will show that there are certain similarities in the dynamical structure of the ergoregions of  naked singularities with  sufficiently high  spins,   {strong naked singularities}, and  black holes with sufficiently small spin,i.e.,  weak black holes, in the sense that none of them
allow any  stable orbiting  matter  inside the ergoregion. In addition, our analysis shows that
an increase of the attractor spin in  weak black hole geometries, or a decrease in strong and   very strong naked singularities, affects  positively the process of formation of a   toroidal orbiting configuration.
We also perform a detailed classification of these sources, taking into account the different dynamical structures  and the peculiarities of the limiting  geometries  defined as the  boundaries of the different classes. We  ultimately conjecture that these peculiar spacetimes could be involved in  spin up and down  processes that lead to a radical change of the dynamical structure of the region closest to the source  and,  therefore,  potentially could give rise to detectable effects.

However it is argued, for example in  \cite{Hawking},  that
the  ergoregion  cannot disappear as a consequence of a spin shift, {as it could be }  filled by negative energy matter {produced by a Penrose process \cite{Penrose71}. In the case of a naked singularity, this  {process} is combined with the repulsive effects of the singularity  ring.}
 Since  a set  of particles with negative energy  could exist in the ergoregion,   one can argue about their  fate   and  role in the source evolution.
These very peculiar  particles differ fundamentally the two types of singularities from the point of view of the  circular motion and their stability:
 the very weak naked singularities  are distinguished by the existence of counterrotating orbits.
The existence of these very orbits can be seen as  a  ``repulsive gravity''  effect  that has been shown to exist also in other spacetimes with naked singularities  \cite{Pu:Charged,Pu:Neutral,Pu:KN}. {This  is   a characteristic   feature  of the  naked singularity geometries \cite{Vie-etal:2014:PHYSR4,Sla-Stu:2005:CLAQG,Kuc-Sla-Stu:2011:JCAP}. In this work, we identify the repulsion effect and its role in the determination of the dynamical structure of the ergoregion  and  the extended matter  in accretion. } This region creates therefore an  ``antigravity''
sphere bounded by orbits with zero angular momentum.
In a bounded region in the antigravity sphere, setting a  bubble of  trapped  (bounded or stable) negative energy particles may form stable or unstable toroidal configurations, explaining, at least for very weak naked singularities, the  fate of particles with negative energies formed according to the Penrose process.
 We evaluate in detail the orbital extension of this region and the  levels of the energies and angular momentum.
 This  material  is obviously formed inside the ergoregion since it cannot penetrate from the external region; the static limit would act  indeed as a {semi-permeable membrane} separating the spacetime  region, filled with negative energy particles, from the external one filled with positive energy particles, gathered from  infinity or expelled from the ergoregion with   impoverishment of the source energy.
In any case, although on the equatorial plane the ergoregion is invariant  with respect to any transformation involving a change in the source spin  (but not with respect to a change in the mass $M$), the dynamical structure  of the ergoregion  is not invariant with respect to a change in the spin-to-mass ratio.
We will show the presence of limiting geometries   where the dynamical structure of the ergoregion outlines a relevant change for a spin shift, where as a consequence of the interaction of the surrounding matter, one would expect {consequently} a shift of the source from one class to another. This effect might have relevant consequences  as a corollary of the runaway instability from relativistic thick disks,  especially for (initial) specific geometries \cite{Abr-Nat-Run,Font:2002bi,Lot2013}. In fact, our analysis enlightens a set of special attractors  and  particularly the spins
${a}_b^-/M\approx0.828427 $ and $a_2/M \approx0.942809$,  where  a slight change in the attractor  spin  produces a {relevant} change in the dynamical structure of the ergoregion, causing  in turn a change in the stability properties of the orbital matter,  and {possibly modifying the closed topology of the  orbiting extended matter configuration which passes from an equilibrium state to the  accretion and, finally, to an  open topology  with matter Jets.}

\medskip
This paper is organized as follows. In Sec. \ref{sec:gen}, we present the Kerr line element and briefly discuss the main physical properties of the corresponding geometry. We present the effective potential that governs the dynamics of the test particle motion on the equatorial plane and, { in Sec.\il\ref{Sec:pro}}, we derive the conditions for the existence and stability of circular orbits. In Secs. \ref{sec:bhs} and \ref{sec:nss}, we explore  the relevant orbital regions in the case of black holes and naked singularities.  In the appendices, we include the definitions of the main radii that determine the dynamical structure of the ergoregion as well as the analysis of the particular limiting cases of extreme black holes and the static boundary.
Finally, in Sec. \ref{sec:con}, we discuss the results and perspectives of our work.

\section{The Kerr geometry}
\label{sec:gen}
The Kerr spacetime is an exact solution of Einstein's equations in vacuum  describing  an axisymmetric,   stationary (nonstatic), asymptotically flat gravitational field.
In spheroidal-like   Boyer--Lindquist  (BL) coordinates
 the Kerr line element has the form
\bea\nonumber && ds^2=-dt^2+\frac{\rho^2}{\Delta}dr^2+\rho^2
d\theta^2+(r^2+a^2)\sin^2\theta
d\phi^2\\\label{alai}&&+
\frac{2M}{\rho^2}r(dt-a\sin^2\theta d\phi)^2\ ,
\eea
where
\be
\Delta\equiv r^2-2Mr+a^2,\quad\mbox{and}\quad\rho^2\equiv r^2+a^2\cos^2\theta \ .
\ee
Here $M$ and $a$  are  arbitrary constants interpreted as  the mass  and rotation parameters, respectively. The specific angular momentum is  $a=J/M$, where $J$ is the
total angular momentum of the gravitational source. In this work, we will
consider the Kerr black hole (\textbf{BH}) case defined by $a\in ]0,M[ $, the extreme black hole source $a=M$, and the naked singularity (\textbf{NS}) case where $a>M$. The  limiting case $a=0$ is the Schwarzschild solution.
The outer and inner \emph{ergosurfaces} $r_{\epsilon}^{\pm}$ and Killing (outer and inner) horizons $r_{\pm}$ are determined respectively by the equations
\(
g_{tt}=0, \ {\rm and} \ g^{rr}=0 \
\)
with
\bea\nonumber&&
r_{\epsilon}^{\pm}\equiv M\pm
\sqrt{M^2-a^2cos^2\theta}\,\quad r_{\pm}\equiv M \pm\sqrt{M^2-a^2},
\\&&\mbox{where}\quad r_{\epsilon}^{-}\leq r_-\leq r_+\leq r_{\epsilon}^{+}\ .
\eea
On the \emph{equatorial plane} $(\theta=\pi/2)$, it is  $\rho=r$ and the spacetime singularity is located at $r=0$ (a curvature ring-like singularity only occurs on
$\theta=\pi/2$ for $M\neq0$, while generally the metric is singular on $\rho=0$, and the space-time is clearly flat for $M=0$).
The region $\Sigma_{\epsilon}^+\equiv]r_+,r_{\epsilon}^{+}[$ for $a\leq M$  (where $g_{tt}>0$) is called {\em ergoregion}.
In black hole spacetimes, the $r$-coordinate is spacelike in the intervals
\(
r\in]0,r_-[  \cup\   r>r_+\ ,
\)
and timelike in the region $r\in]r_-,r_+[$. This means that the surfaces of constant $r$, say\footnote{{$\Sigma_{\mathbf{Q}}$ is the   $\mathbf{Q}=$constant surface for any quantity or set of quantities $\mathbf{Q}$.}} $\Sigma_r$,  are timelike for $\Delta>0$, spacelike for $\Delta<0$ and null for $\Delta=0$.
On the other hand,  for  $r\in]r_{\epsilon}^{-},r_{\epsilon}^{+}[$ the metric component $g_{tt}$ changes its sign and vanishes for $r=r_{\epsilon}^{\pm}$ and $\cos^2\theta\in]0,1]$, and also at $r=2M$
for $\theta=\pi/2$.
As the spin of the  attractor   reaches the limiting value $a=M$ of  an extreme \textbf{BH} source, the horizons coincide, $r_-=r_+ =M$, and
if $\cos^2\theta=1$, i.e., on the rotational axis, it is
$
r_{\epsilon}^{\pm}=r_{\pm}$. 
%
%
%
On the other hand, in naked singularity geometries,  the radii $r_{\pm}$ are not real and the singularity at $\rho=0$ is not covered by a  horizon.  However,   the ergosurfaces $r_{\epsilon}^{\pm}$ are still  well defined\footnote{In particular it is
\(
{ g_{tt}<0}\)
in $0\leq \cos\theta^2\leq{M^2}/{a^2}$ for
$r\in]0,r_{\epsilon}^-[\cup r>r_{\epsilon}^+$ and
in $\cos\theta^2\in]{M^2}/{a^2}, 1]$ for $r>0 $.
}.
We can define as well on the equatorial plane the ergoregion $\Sigma_{\epsilon}^+\equiv]0,r_{\epsilon}^{+}[$ for $a> M$
and $\Sigma_{\epsilon}^+$ has a toroidal topology   centered on the  axis whose inner
circle is the naked  singularity.
In this work, we focus on  the dynamics inside the ergoregion on  $\theta=\pi/2$,  where { $\left. r_{\epsilon}^+\right|_{\pi/2}=\left.r_+\right|_{a=0}=2M$} and $r_{\epsilon}^-=0$,
   and we compare   the dynamical properties of circularly orbiting matter in $\Sigma_{\epsilon}^+$  for the \textbf{BH} and \textbf{NS} cases.

{The motion of matter in    $\Sigma_{\epsilon}^+$  has very peculiar properties. We summarize here some  of the major characteristics  relevant  for this analysis. First,}
the  equatorial (circular) trajectories   are confined in the equatorial geodesic plane as a consequence of the metric tensor  symmetry under reflection through  the  equatorial hyperplane $\theta=\pi/2$.
A fundamental property of the ergoregion from the point of view of the matter and field
dynamics   is that  no matter can be at rest in $\Sigma_{\epsilon}^+$  (as seen  by a faraway observer or, in other words, from infinity in a BL coordinate frame) \cite{Landau}. Then, the metric is no longer stationary.
As we have already noted,  in the BL coordinates   the surfaces of constant $(r, \theta, \phi)$, {with   line element}
 \(\left.ds\right|_{\Sigma_{r,\theta, \phi}}\),  are  spacelike  inside the ergoregion,
 that is, the ``time'' interval  becomes spacelike, and in terms of BL coordinates this means that $t$  is spacelike and  any motion  projected into $\Sigma_{r,\theta, \phi}$ is forbidden.
The outer ergosurface $r_{\epsilon}^+$  is called the static surface or static limit (also stationary limit surface, see \cite{GriPod09}).
This is   a timelike
surface except on the axis of a Kerr source where it matches the outer horizon and then  it is null-like.  At $r>r_{\epsilon}^+$  particles  can follow an   orbit of the vector $\xi_t$ and eventually cross the static limit (away from the axis).
Moreover,  for a timelike particle (with positive energy) it is  possible to cross the static limit and
to escape towards infinity.
In the  static \textbf{BH} spacetime  ($a=0$), the region   $]0,\left.r_+\right|_{a=0}[$  coincides with the zone inside the horizon; then, no particle can stay at rest (with respect to an observer located at infinity) neither at $r=$constant, i.e., any particle is forced to fall down into the singularity.
For the stationary spacetimes ($a\neq0$)     in $\Sigma_{\epsilon}^+$ the motion with $\phi=const$ is \emph{not} possible and all particles are forced to rotate with the source i.e.  $\dot{\phi}a>0$. Indeed, in the  ergoregion  the Killing vector $\xi_t^{a} = (1, 0, 0, 0)$ becomes spacelike, i.e.,  {$g_{ab}\xi_t^a\xi_t^b=g_{tt}>0$}.
This fact implies in particular that a  {\em static observer},  i.e. an observer with  four-velocity  proportional
to $\xi_t^a$ so that $\dot{\theta}=\dot{r}=\dot{\phi}=0$, {(the dot denotes  the derivative with respect to the  proper
time $\tau$ along the curve)},
cannot exist inside
the ergoregion.
Therefore for any in-falling matter (timelike or photonlike) approaching  the horizon $r_+$, in the region $\Sigma_{\epsilon}^+$, it is $t\rightarrow\infty$ and $\phi\rightarrow\infty$ meaning that the world-lines  around the horizon, as long as $a\neq0$, are subject to  an
 infinite twisting.
Trajectories  with $r=const$ and $\dot{r}>0$ (particles  crossing the static limit and escaping outside  in $r\geq r_{\epsilon}^+$) are possible.
 Another important point is that for an observer at infinity, the particle  will reach and penetrate the  surface $r=r_{\epsilon}^+$,
   in general, in a finite time $t$. For this reason, the ergoregion boundary is not a surface of infinite redshift, except for the axis of rotation where the ergoregion coincides with the event horizon. Indeed, concerning the frequency of a signal emitted by a source in motion along the boundary of the ergoregion $r_{\epsilon}^+$, it is clear that the proper time of the source particle
is not null \footnote{However, as it is  $g_{tt}(r_{\epsilon}^{\pm})=0$ in is also named infinity redshift surface see for example \cite{GriPod09}}.
This means that the observer at infinity
will see a non-zero emission frequency. In the spherical symmetric case $(a=0)$, however, as $g_{t\phi}=0$
the  proper time interval $d\tau=\sqrt{|g_{tt}|}dt$,  goes to zero as one approaches $r=r_+=r_{\epsilon}^+$.
Thus, on the equatorial plane as $a\rightarrow 0$ and the geometry ``smoothly" resembles the spherical symmetric case the frequency of the emitted signals, as seen by an observer at infinity, goes to zero.
%
\subsection{On the particle's energy and the  effective potential}\label{Sec:pro}
In this section, we discuss some aspects of   particle energy definition and the effective potential for the circular orbits, which will be  used here to analyze the dynamics inside the ergoregion $\Sigma_{\epsilon}^+$.
Let $ u^{a}={dx^{a}}/{d\tau}=\dot{x}^{a}$ be the tangent vector to  a curve $x^a(\tau)$,
the  momentum $p^a= \mu\dot x^a$ of a particle with  mass $\mu$
is normalized so that
$g_{ab}\dot{x}^{a}\dot{x}^{b}=-k$, where {$k=0,-1,1$} for null, spacelike and timelike curves, respectively.
In order to simplify the  investigation of the circular dynamics we  use  the   symmetries  of the Kerr geometry:
since the metric tensor is independent of $\phi$ and $t$, the covariant
components $p_{\phi}$ and $p_{t}$ of the particle's four-momentum are
conserved along its geodesic, i.e., the quantities
\bea\nonumber
E& \equiv& -g_{ab}\xi_{t}^{a} p^{b}=-\left(g_{tt}p^t+g_{t\phi}p^{\phi}\right),
\\\label{Eq:E}
 L &\equiv&
g_{ab}\xi_{\phi}^{a}p^{b}=g_{\phi\phi}p^{\phi}+g_{t\phi}p^{t}
\eea
are  constants of motion, where  $\xi_{\phi}=\partial_{\phi} $ is the
rotational Killing field   and
$L$ is interpreted as the angular momentum  of the particle as measured by an observer at infinity.  The  Killing field $\xi_{t}=\partial_{t} $ represents
 the  stationarity of the spacetime and
we may interpret $E$, for
timelike geodesics, as  the total energy of the test particle
coming from radial infinity, as measured  by  a static observer located  at infinity.

{An essential characteristic of the region  we are investigating  is to foreseen negative energy states for the  matter dynamics, therefore   it is convenient here to give a more detailed definition of the  energy.}
In general, the particle's energy could  be defined   as {${E}_{\tau}$ in terms of
$
{\partial \mathcal{S}}/{\partial \tau}$  or  as ${E}_{t}\equiv E$ in terms of ${\partial \mathcal{S}}/{\partial t},
$ }
where  $\mathcal{S}$  is the particle action. The energy  ${E}_{\tau}$ is defined with respect to the proper time of the particle synchronized along the trajectory. This quantity is always positive but  in general not  conserved.
On the other hand, the definition of ${E}_{t}$  contains the derivative with respect to the universal time, taking account of the symmetries of the stationary spacetime, and consequently it  is constant along the orbit of the timelike Killing vector. This quantity, as defined in Eq.\il(\ref{Eq:E}), is conserved, but can be negative in $\Sigma_{\epsilon}^+$, where $t$ is no more a timelike coordinate
 \cite{Landau}. The particle with negative energy cannot escape to infinity, but its dynamics
is   confined in $r<r_{\epsilon}^+$.
Therefore the static limit would act  as a \emph{semi-permeable membrane} separating the spacetime  region  $\Sigma_{\epsilon}^+$, filled with negative energy particles, from the external one filled with  particles with positive energy, gathered from   infinity or expelled from the ergoregion with the consequent impoverishment of source energy. However, in the case of a naked singularity this phenomenon is combined with the repulsive effects from the singularity  ring $r=0$,  and consequently the region $\Sigma_{\epsilon}^+$ will have a more articulated  structure of the  matter dynamics. {Possibly, as pointed out in \cite{Hawking},  a  set  of negative energy particles   will form  (from a Penrose process) in $\Sigma_{\epsilon}^+$,  so  one task is to argue  what is the fate of the negative energy particles and what role can these particles play   in the evolution of the source.}
The relevance of the speculated Penrose process   in the eventual formation  in $\Sigma_{\epsilon}^+$ of an extended configuration of  negative energy matter cannot be ruled  by considerations of the geometric properties of spacetime only. However,
in this work  we address the question of what would happen to the particle dynamics in $\Sigma_{\epsilon}^+$ as well as to the dynamics of the  extended  object  under hydrostatic pressure, focusing in particular  on the energetic properties of the test  particles  and on the necessary conditions for the  formation and stability of toroidal, pressure supported accretion disks. 
As  we focus  here on the specific case of circular motion,  one could ask  if  the circular motion (stable or not) of particles with negative energy is  possible for some kind of sources.

{We start our analysis noting that, }
using the definitions given  in Eq.\il(\ref{Eq:E}),
the investigation of the circular  motion of test particles in the equatorial plane can be
 reduced to the study of  the motion in the effective potential $V$,
defined from the {\em normalization condition} of the particle's  four-velocity
\bea\label{Eq:norm-co}
g_{tt} \dot{t}^2+ g_{\phi\phi}\dot{\phi} ^2+2g_{\phi t} \dot{t}  \dot{\phi} +g_{rr} \dot{r}^2=-k \ .
\eea
where the condition  $\dot{\theta}=0$ has been used as geodesics starting
in the equatorial plane are planar.
Using Eqs.\il(\ref{Eq:E})
for a particle  in circular motion, i.e. $\dot{r}=0$,  we obtain from Eq.\il(\ref{Eq:norm-co}) the effective potential
\bea
V^{\pm}=\frac{-g_{\phi t} L\pm\sqrt{ \left(g_{\phi t}^2-g_{tt} g_{\phi\phi}\right) \left(L^2+g_{\phi\phi} k\mu^2\right)}}{g_{\phi\phi}},
\eea
which represents the value of $E/\mu$ that makes $r$ into a ``turning
point'' $(V=E/\mu)$; in other words, it is the value of $E/\mu$ at which
the (radial) kinetic energy of the particle vanishes.
The (positive) effective potential can be written explicitly as
\begin{equation}\label{qaz}
V\equiv  \frac{V^{+}}{\mu}=-\frac{\beta}{2\alpha}+\frac{\sqrt{\beta^2-4 \alpha \gamma}}{2\alpha},
\end{equation}
where \cite{MTW,RuRR}
\bea\nonumber
\alpha&\equiv&\left(r^2+a^2\right)^2-a^2\Delta,\quad
\beta\equiv-2aL\left(r^2+a^2-\Delta
\right),\\
\gamma&\equiv&a^2L^2-\left(M^2r^2+L^2\right)\Delta\ .
\eea
The behavior of the  effective potential  $V^-$
can be studied by using the following symmetry
\(
V^{+}(L)=-V^{-}(-L)
\).
The Kerr metric (\ref{alai}) is invariant under the application of any two different transformations:
\(\mathbf{\mathcal{P}}_{\mathbf{Q}}:\mathbf{Q}\rightarrow-\mathbf{Q},
\)
where $\mathbf{Q}$  is one of the coordinates $(t,\phi)$ or the metric parameter $a$, a single transformation
leads to a spacetime with an opposite  rotation respect to the unchanged metric. Thus, we fix $a\geq0$
and, noting that the potential function (\ref{qaz}) is invariant under the mutual transformation of the parameters
$(a,L)\rightarrow(-a,-L)$. We will restrict the  analysis to the case of  positive values of $a$
for corotating  $(L>0)$ and counterrotating   $(L<0)$ orbits.
For (timelike) circular orbits it is
\be\label{Eq:Kerrorbit}
\dot{r}=0,\quad V =\frac{E}{\mu},\quad \frac{d V }{d r}=V'=0
\ee
 (see also \cite{Pu:Kerr}). Solving  the equation $V'=0$ with respect to the angular momentum, we find
\be\label{LPM}
\frac{L_{\pm}}{\mu M}\equiv\frac{\left|\frac{a^2}{M^2}\pm
2\frac{a}{M}
\sqrt{\frac{r}{M}}+\frac{r^2}{M^2}\right|}{\sqrt{\frac{r^2}{M^2}\left(\frac{r}{M}-3\right)\mp
2\frac{a}{M}\sqrt{\frac{r^{3}}{M^3}}}} \ .
 \ee
Any particle moving along a circular orbit in these spacetimes has  an angular momentum {of magnitude} either $L_+\geq0$ or $L_-\geq0$. Introducing  Eq.\il(\ref{LPM}) into (\ref{qaz}), we find  the   energies
$
{{E}_{\pm}}\equiv{{E}(L_{\pm})}$ and  $
{{E}^{-}_{\pm}}\equiv{{E}(-L_{\pm})}
$.
The orbits with angular momentum $L=\mp L_{\pm}$ are allowed in different orbital regions  for different attractors, depending on the corotating or counterrotating nature of the motion.

The energies $\mathcal{E}\equiv\{{{E}_{\pm}},{{E}^{-}_{\pm}}\}$ and the angular momenta $\mathcal{L}\equiv \{L_{\pm},-L_{\pm}\}$ are functions of the spacetime spin-mass ratio and, therefore,  we   use them in this work to characterize different attractors with different  spin-mass ratios.
In general, depending on the context, under   $\mathcal{E}$ or $\mathcal{L}$ we shall understand, if no otherwise specified, either the set of the respective quantities or a generic element of the set, which will be specified when necessary.
The boundaries of these regions are determined by a set $r_i\in\mathcal{{R}}$,  different for  \textbf{BH} and \textbf{NS} configurations and includes the radii $\{r_{\gamma},r_{b},r_{lsco}\}$ as given in Table\il\ref{Table:asterisco}. These radii are  defined through different expressions   for corotating $(-)$ and counterrotating orbits $(+)$  in    naked singularity or  black hole geometries respectively {(see also Sec.\il\ref{Sec:po})}. The physical importance of this radii can be explained as follows:   Timelike  circular orbits  can fill  the spacetime region $r>r_{\gamma}$, and the   orbits   at $r_{\gamma}$, as defined   in Eq.\il(\ref{Eq:gamma}), are photon orbits (light-like orbits), known as marginally circular orbits  or also \emph{last circular orbits}.  It should be noted, however, that  orbits with radius $r_{\gamma}$ are \emph{not} timelike, but  they are limiting values for timelike free particles and matter where the  {angular momentum}  diverges.
Thus no  circularly orbiting (timelike)  matter can be formed in the region {$\Sigma_{\varnothing}\equiv ] r_+ r_{\gamma}]$}, because the energy and angular momentum of the particle diverge as the photon-like orbit is approached.

Stable orbits are in the region $\Sigma_{s}\equiv]r_{lsco},+\infty[$, and the orbits at $r_{lsco}$,  defined in Eqs.\il(\ref{Eq:rlscomp},\ref{Eq:rlscompNS}), are named, accordingly, \emph{ marginally stable circular orbit} or also, quite improperly, \emph{last} stable circular orbits (see, for example, \cite{Wald74,Wald}), {respectively for \textbf{BH} and \textbf{NS} sources and for different types of orbits, as specified in  Sec.\il\ref{Sec:po}}. While it should be kept in mind that $r_{lsco}$ does  \emph{not} correspond to a stable orbit, as it is indeed a saddle point for the  effective  potential, that is $\left. d_rV\right|_{r_{lsco}}=\left.d^2_rV\right|_{r_{lsco}}=0$, the marginally stable orbits correspond to the minimum of the energy and angular momentum of the particle \cite{Pu:Kerr}).

The instability region  $\Sigma_{u}\equiv]r_{\gamma},r_{lsco}]$   is  split by the radius
 $r_{b}$, where $\mathcal{E}(r_b)=1$, generally into two regions $\Sigma_{u}=\Sigma^{\geq}_{u}\cup\Sigma^<_{u}$ where
$r_{b}$, defined in Eqs.\il(\ref{Eq:perd-BH},\ref{Eq:perd-NS}), determines  the
 \emph{marginally  bounded orbit} or \emph{last} bounded orbit. In $\Sigma^{\geq}_{u}\equiv]r_{\gamma},r_{b}]$ it is $\mathcal{E}\geq1$ and in $\Sigma^{<}_{u}\equiv]r_{b},r_{lsco}]$ it is $\mathcal{E}<1$. {In Sec.\il\ref{Sec:firstNS}, we will discuss the special  case where the energy parameter can be  negative.}
The explicit  expressions of the radii $\mathcal{R}$ for black holes are given, for example, in \cite{Pu:Kerr}. In this work, we focus on the case of naked singularities, specifying the exact form of $r_i\in\mathcal{{R}}$ for different classes of super-spinning  objects ($a>M$).

The parameters
$\mathbf{p}\equiv\{\mathcal{L},\mathcal{E}\}$ and  the radii $\mathcal{R}$ determine  the main properties of  test particles moving along circular orbits. These are the main quantities that will be used below to explore the physical properties of spacetimes described by the Kerr metric.
  We investigate the dynamics  in the ergoregion $\Sigma_{\epsilon}^+$ through the study of the regions $\Sigma\equiv\{\Sigma_{s},\Sigma_u, \Sigma_{\varnothing}\}$, which are characterized by the geometric properties
	$\Sigma_{\epsilon}^+=\bigcup\Sigma_i$, and $ \Sigma_j\bigcap\Sigma_i=\emptyset,\quad\forall i,j$. The dynamical
  structure of the ergoregion is important because it
   fully characterizes the dynamical  properties of the orbiting matter inside   $\Sigma_{\epsilon}^+$.

The properties of matter in circular configurations, as well as  toroidal configurations  of extended matter in accretion disks,  are {in general}  governed by several factors such as   pressure,   magnetic fields,  dissipative effects and so on; however, the properties of the spacetime structure,
regulated by the radii $\mathcal{R}$,   are the   basis on which to build any disk model \cite{Abramowicz:2011xu},
 especially in thick pressure supported   relativistic accretion disks where the effects of the background curvature   strongly affects  the equilibrium and the disk instability.
The disk equatorial plane of symmetry is often aligned with the symmetric plane of the accreitor. By using the symmetry of the system, the geometric symmetries  of the background and of the matter  configuration, we are capable to reduce  the  analysis to an one-dimensional problem along the radial direction
$r$,  the dynamics of the orbiting matter   being regulated entirely by the couple of  model parameters $\mathbf{p}$, which are constant  along the  geodesics because they are associated with the  orbits of the  Killing vector fields.
Then the main features of the disk dynamics, stability and accretion, are mainly regulated by the properties of the disk as  projected on the plane of symmetry $\theta=\pi/2$.
Moreover, the region $\Sigma_{\epsilon}^+$  is closely involved in the evolutionary processes of the accreitor, resulting in a shift of the intrinsic spin. As a consequence of the interaction of the surrounding matter in $\Sigma_{\epsilon}^+$, one would expect a change of the source spin, resulting in a change of the corresponding class or subclass {of the attractor}, with potentially relevant phenomenological implications
\cite{Abr-Nat-Run,Font:2002bi,Lot2013}.
Furthermore, in \cite{Hawking} it is argued that
the  ergoregion (in any plane $\theta$) cannot disappear as a consequence of a spin shift as it is supposed to be filled by the negative energy matter;
in fact, the limiting case corresponds to $r_+=r_{\epsilon}^+=2M$.
On the equatorial plane, the ergoregion (for $a\neq0$) is invariant with respect to any transformation involving a change in the source spin  (but not with respect to  a change in the mass $M$), but the dynamical structure  of $\Sigma_{\epsilon}^+$ is not  invariant under changes of the ratio $a/M$; in other words, the limit  $\left.r_{\epsilon}^+/M\right|_{\theta=\pi/2}= 2$  is independent of the  central attractor,  but not its {dynamical} structure  $\Sigma$.
We will show the presence of limiting geometries from the point of view of the structure of  $\Sigma_{\epsilon}^+$, where the dynamical structure of the ergoregion {can change significantly as a consequence of a small shift of the attractor spin } (see also  Sec.\il\ref{sec:ebh}).

\section{Analysis and discussion of the dynamics}
\label{Sec:listdis}
In this section, we discuss the   particle dynamics  circularly orbiting in  $\Sigma_{\epsilon}^+$,  and characterizing the   different attractors in terms of the   dynamical structure  $\Sigma$  of the ergoregion.  The case of $\mathbf{BH}$ geometries is addressed in Sec.\il\ref{sec:bhs},  and the \textbf{NS} case is considered in Sec.\il\ref{sec:nss}. A similar  analysis for  the region $r>r_{\epsilon}^+$ was discussed  in     \cite{Pu:Kerr}.   Our results   lead to the  identification of  different classes of attractors, defined by  the spin-mass ratios in the   regions of the $r-a$ plane    with boundaries in  $\mathcal{R}\otimes\mathcal{A}$, where $\mathcal{A}$ is the set of spins defined  at the crossings of the radii $(r_{i},r_j)$ in $\mathcal{R}$  and $r_{\epsilon}^+$ in $\Sigma_{\epsilon}^+$
(see Table\il\ref{Table:asterisco}).
We identify four  classes of black hole  sources $(a\leq M)$ and five classes of naked singularities $(a>M)$   defined in
Table\il\ref{Table:asterisco}. The  classes of geometries have  boundaries in  $\mathcal{A}=\mathcal{A}_{BH}$ and $\mathcal{R}=\mathcal{R}_{BH}$ for black holes, and  $\mathcal{A}=\mathcal{A}_{NS}$ and $\mathcal{R}=\mathcal{R}_{NS}$ for naked singularities (details are given in Table\il\ref{Table:asterisco}).
The geometries with spin $a_i\in\mathcal{A}_{BH}\cup\mathcal{A}_{NS}$ are limiting cases with respect to  the structure of $\Sigma_{\epsilon}^+$. Particularly relevant is the extreme case $a=M$ which marks also the limit between black holes and naked singularities; this case will be addressed in particular in Sec.\il\ref{sec:ebh}. Close to the boundary spacetimes, { in particular, as discussed in Sec.\il\ref{sec:ebh}, at $a=a_2\approx 0.942809 M$ and $a=M$,} the properties  around the limiting spin value are rather subject to a sort of fine-tuning, i.e. at a fixed radial distance from the source, the dynamical properties modify   significantly as consequence of a  slight change in the  spin (see Sec.\il\ref{sec:ebh}). For this reason, the geometries with the spin in $\mathcal{A}$ are likely to give rise to rather relevant  phenomena for the geometrical properties of the  matter  configurations  orbiting in regions very close to the source, {see also Sec.\il\ref{sec:ebh}}.

Photon-like circular orbits in $\Sigma_{\epsilon}^+$ are  a feature of  \textbf{BHII} and \textbf{BHIII} sources only; there are no photon-like orbits inside the ergoregion  $\Sigma_{\epsilon}^+$ in   \textbf{NS}-spacetimes. In   \textbf{NS}-spacetimes, there is no last circular orbit for  corotating particles with $L=L_-$, indicating that  circular orbits  can theoretically be up to the singularity. In other words, in this case there is at least one circular orbit for any  $r>0$ as long as the particle's angular momentum  assumes certain values. On the other hand,  counterrotating orbits $(L=-L_-)$,  characteristic of   \textbf{NSI} sources, can exist only in  a bounded orbital region.
Stability  regions $\Sigma_{s}\subset\Sigma_{\epsilon}^+$ are present only in classes \textbf{BHIII}, \textbf{NSI} and \textbf{NSII}
\begin{table*}
\centering
\caption{Set of spins $\mathcal{A}$ and radii $\mathcal{R}$ for  the black hole (\textbf{BH}) and the naked singularity (\textbf{NS}) geometries, respectively.  The radii $\mathcal{R}_{BH}=\{r_{\gamma}^-, r_b^-,r^-_{lsco}\}$ set the photon orbit (or also last circular orbit), the marginally bounded orbit, and the  last stable circular orbit, respectively (for corotating orbits)  in black hole geometries. See also Figs.\il\ref{rewpoett} and \ref{rewpoettx}, and the discussions in Sec.\il\ref{sec:bhs} for the  \textbf{BH} case, and Sec.\il\ref{sec:nss} for the \textbf{NS}  case. For the naked singularity case the radii $\mathcal{R}_{NS}=\{r_{\upsilon}^{\pm},\hat{r}_{\pm},r_{lsco}^{(NS)-},r_b^{(NS)}\}$ set the outer $(r_{\upsilon}^+)$ and inner $(r_{\upsilon}^-)$ {effective ergosurfaces}, where the orbital energies ${E}=0$ (see  Sec.\il\ref{sec:nss}), and  the radii $\hat{r}_{\pm}$ ({zero angular momentum radii}),  where  $\mathcal{L}(\hat{r}_{\pm})=0$. Moreover, $r_b^{(NS)}$  and the $r_{lsco}^{(NS)-}$ are the marginally bounded orbit and the last stable circular orbit,  respectively, for \textbf{NS} geometries.
The explicit expression of the radii can be found in  Sec.\il\ref{Sec:po}}
\label{Table:asterisco}
\begin{tabular}{lrcl}
\hline
\textbf{Black hole  classes:} $\mathcal{A}=\mathcal{A}_{BH}\equiv\{0,a_1,a^-_{b},a_2,M\}$, $\mathcal{R}=\mathcal{R}_{BH}\equiv\{r_{\gamma}^-,r_{b}^-,r_{lsco}^-\}$
\\\\
${a}_1/M\equiv1/\sqrt{2}\approx0.707107;\quad{a}_b^-/M\equiv 2 (\sqrt{2}-1)\approx0.828427 ;\;\quad a_2/M\equiv {2 \sqrt{2}}/{3}\approx0.942809$
\\\\
$a_1:\;r_{\gamma}^-(a_1)=r_{\epsilon}^+;\quad a_b^-:\;r_{b}^-({a}_b^-)=r_{\epsilon}^+;\quad a_2:\; r_{lsco}^-(a_2)=r_{\epsilon}^+$
\\\\
$\mathbf{BHI}:\; [0,a_1[\quad \mathbf{BHIIa}:\; [a_1,a_b^-[;\quad \mathbf{BHIIb}:\; [a_b^-,a_2[;\quad \mathbf{BHIII}:\;[a_2,M]$
\\\\
 \hline 
\textbf{Naked singularity classes:}
$\mathcal{A}=\mathcal{A}_{NS}\equiv\{a_{\mu}, a_3,a_4, a_{b}^{NS}\}$,\quad $\mathcal{R}=\mathcal{R}_{NS}\equiv\{r_{\upsilon}^{\pm},\hat{r}_{\pm},r_{lsco}^{(NS)-},r_b^{(NS)}
\}$
\\ \\
$a_{\mu}/M={4 \sqrt{{2}/{3}}}/{3}\approx1.08866; \quad{a}_3/M\equiv 3\sqrt{3}/4\approx1.29904;\;\quad
a_4/M\equiv 2 \sqrt{2}\approx2.82843:\;\quad a_b^{NS}/M\approx 4.82843$
\\\\
$a_{\mu}:\;r_{\upsilon}^+({a}_{\mu})=r_{\upsilon}^-({a}_{\mu});
\quad a_3:\;\hat{r}_{+}(a_3)=\hat{r}_{-}(a_3);
\quad a_4:\;r_{lsco}^{(NS)-}(a_4)=r_{\epsilon}^+;\quad a_b^{NS}:\; r_{b}^{(NS)}(a_b^{NS})=r_{\epsilon}^+$
\\\\
$\mathbf{NSIa}:\; ]M,a_{\mu}];\quad  \mathbf{NSIb}:\; ]a_{\mu},a_{3}];\quad \mathbf{NSII}:  ]a_3, a_4];\quad \mathbf{NSIIIa}:\; ]a_4, a_b^{NS}];\quad \mathbf{NSIIIb}:\; ]a_b^{NS},+\infty]$
\\\\
\hline
\end{tabular}
\end{table*}
(see Table\il\ref{Table:asterisco} and Figs.\il\ref{rewpoett} and \ref{rewpoettx}).

We will investigate the {structure}  $\Sigma$ of   $\Sigma_{\epsilon}^+$, providing a  characterization of the regions of $\Sigma$ and,
in particular, the length or  extension  of each orbital region in which $ \Sigma_{\epsilon}^+$ is decomposed,  where
the extension $\mu_{\epsilon}^+=2M$ of $\Sigma_{\epsilon}^+$ is  the sum of the lengths of all regions  of $\Sigma$.
To understand the physical implications of this analysis, we will focus our discussion on two  different orbiting  configurations:  the case of  Keplerian (dust) disks and the  pressure supported accretion thick disk models \cite{Abramowicz:2011xu,Sla-Stu:2005:CLAQG,Adamek:2013dza,Pugliese:2014bua,PuMonBe12,Joshi:2013dva}.
We  discuss the  physical meaning  of the regions  $\Sigma_s$ $\Sigma_u$ and $\Sigma_{\varnothing}$ in relation to these astrophysical models.

The region  $\Sigma_{s}$,
 consists of  stable geodesic orbits and, in the case of a dust disk, the length $\mu_s$ of $\Sigma_s$ indicates the amount of dust that can be in equilibrium inside the ergoregion $\Sigma_{\epsilon}^+$.
According to the current  accretion disk models, the upper bound for the  inner edge of the disk is actually located  in a region $ r\leq r_{lsco}$ then,
the length $ \mu_s$ of the orbital region $\Sigma_s\subseteq \Sigma_{\epsilon}^+$  provides the maximum elongation, on the equatorial plane inside the ergoregion, of a hypothetical accretion disk  with inner edge   in  $\Sigma_{\epsilon}^+$. Then, the disk  penetrates the ergoregion, crossing  the static limit (see, for example, \cite{Pugliese:2014bua}), or it is  totally included in  $\Sigma_{\epsilon}^+$.
If the disk is thick, one can use the Boyer model (see \cite{Boy:1965:PCPS:}) for a  pressure supported thick accretion disk, where the  center of the configuration  $r_c$, which locates the maximum  of the hydrostatic pressure, is  inside the stable region, i. e., $r_c\in\Sigma_s\subseteq\Sigma_{\epsilon}^+$.
However, the inner edge $r^i_{\mathcal{K}}$ of the disk, in the case of thick disk in equilibrium as established by the Paczynski-Wiita (P-W)  mechanism \cite{Abramowicz:2011xu}, can be also located in a region $r_b<r_{max}<r^i_{\mathcal{K}}<r_{lsco}<r_{min}<r^o_{\mathcal{K}}$, where $r^o_{\mathcal{K}}$ denotes the outer edge and $(r_{max},r_{min})$ are  the  maximum and minimum  points of the effective potential, respectively.

The instability region   $\Sigma_{u}$, filled with  unstable circular orbits, is generally split  into two regions (that  may be connected or not) as $\Sigma_{u}=\Sigma_{u}^{\geq}\cup\Sigma_{u}^{<}$.
In the  instability region,  {there could be a}  decay phenomenon  in which a particle may either escape, spiraling into the outer region and therefore could become observable, or be captured by the source changing its spin-to-mass ratio (see, for example, \cite{MTW}). This phenomenon, in turn, could also may give rise to Jets of matter.
The instability region   has indeed  an important role for the relativistic thick accretion disk model  as it is essential for the modelization  of associated
non-equilibrium phenomena related either to the accretion or to the Jet production.
When the  inner edge of the disk  is in $r^i_{\mathcal{K}}\in\Sigma_{u}^{<}$  (on a maximum point of the effective potential) while  $r_{c}\in\Sigma_{s} $, {can give}  rise to a limiting situation, then $r^i_{\mathcal{K}}$ represents an unstable point, a cusp in which matter grows on the attractor or, also  in some geometries, a point of  excretion outwardly. In any case, the material presents a lobe closed to and centered around the maximum pressure point $r_c$, while the cusp point represents the overflow point of the material.
Then the length $\mu_s+\mu_{u}^{<}$, where  $\mu_{u}^{<}$ is the length of $\Sigma_{u}^{<}$, provides the maximum elongation on the equatorial plane of the  distance  $r_c-r^i_{\mathcal{K}}$ for an accretion disk in  $\Sigma_{\epsilon}^+$.
If the disk is thick,  then the density is constant and one could  evaluate, in terms of the length $\mu_s+\mu_{u}^{<}$, the total mass  contained in the configuration with maximum elongation.
 In  $ \Sigma_{u}^{\geq}$, where $\mathcal{E}\geq1$,  circular geodesics  are unstable  and unbounded.
According with the Boyer model \cite{Boy:1965:PCPS:}, the extended matter  configurations  with   a minimum point of pressure located in $ \Sigma_{u}^{\geq}$ {can} open in  Jets  with funnels along the axis.
More precisely, the maximum extension $\lambda$  for a pressure supported  thick  accretion disk  (see, for example, \cite{Abramowicz:2011xu}) is the elongation $\lambda_x$ on the equatorial plane of its critical configuration, i.e., in accretion.  Then,
  the inner edge  $r_{\mathcal{K}}^i$ is located exactly at the maximum of the effective potential and precisely
 $r_{\mathcal{K}}^i=r_{max}\in \Sigma_{u}^{<}$ and  the outer edge is also uniquely fixed  by $ r_{max} $. For  sufficiently large angular momenta,   such that the disk is entirely included in $\Sigma_{\epsilon}^+$,  then
 the upper bound of its elongation $\lambda\equiv r_{\mathcal{K}}^o-r_{\mathcal{K}}^i$
is $\sup{\lambda_x}= r_{\mathcal{K}_x}^o-r_{\mathcal{K}_x}^i\approx\mu_{u}^{<}+r_{\epsilon}^+-r_b$ (this is because we are are considering in this work the region $r<r_{\epsilon}^+$ only  and   $r_{lsco}>r_{b}>r_{\gamma}$;  we could equivalently say that $\sup{\lambda}=\mu_{u}^{<}+\mu_s$).
The elongation of the disk in equilibrium, centered in $r_c=r_{min}$, must  be lower than this critical value and   its inner  and outer edges are included  in the range  $]r_{\mathcal{K}}^i,r_{\mathcal{K}}^o[\subset]r_{\mathcal{K}_x}^i,r_{\mathcal{K}_x}^o[$. Then, for example, in  $\mathbf{BHIII}$ spacetimes   it is  $\sup{\lambda}\leq M$ (recall that in  \textbf{BHIII} spacetime it is $\partial_a\mu_s>0$). However,  even if the evaluation of $\lambda$ depends on the adopted angular momentum, for the equilibrium configuration it is
$r_{max}< r_{lsco}<r_{min}=r_c<r_{\mathcal{K}}^o$
while   for the accretion configuration it is
$r_{max}=r_{\mathcal{K}_x}^i<r_{lsco}<r_{min}<r_{\mathcal{K}}^o<r_{\mathcal{K}_x}^o$
 with $r_{\mathcal{K}_x}^i<r_{\mathcal{K}}^i$ and thus, for instance, in the  case of \textbf{BHIIb} sources,
where the accretion of even Jets can occur,  the supremum of the distance $r_c-r_{\mathcal{K}_x}^i$ is  $\sup{\mu_{u}^{<}}$.

In general,
one might look at the entire evolution of a thick accretion disk, from the formation of a thin ring, its consequent growing  to the unstable phases of accretion which results finally in the formation of Jets of matter from the cusp point, as the transition of orbiting matter  through the regions $\Sigma_{u}\rightarrow\Sigma_{u}^<\rightarrow\Sigma_{u}^{\geq}$. {Finally}, the analysis of the dynamics in the region $\Sigma_{\epsilon}^+$ allows us to consider, from the starting condition, the  extraction of energy and angular momentum of a black hole with  the consequent release  of particles (or matter Jets), that could eventually induce a shift of the attractor from one class to another.
\subsection{Black holes}\label{sec:bhs}
\begin{figure}
\begin{tabular}{cc}
\includegraphics[scale=0.3]{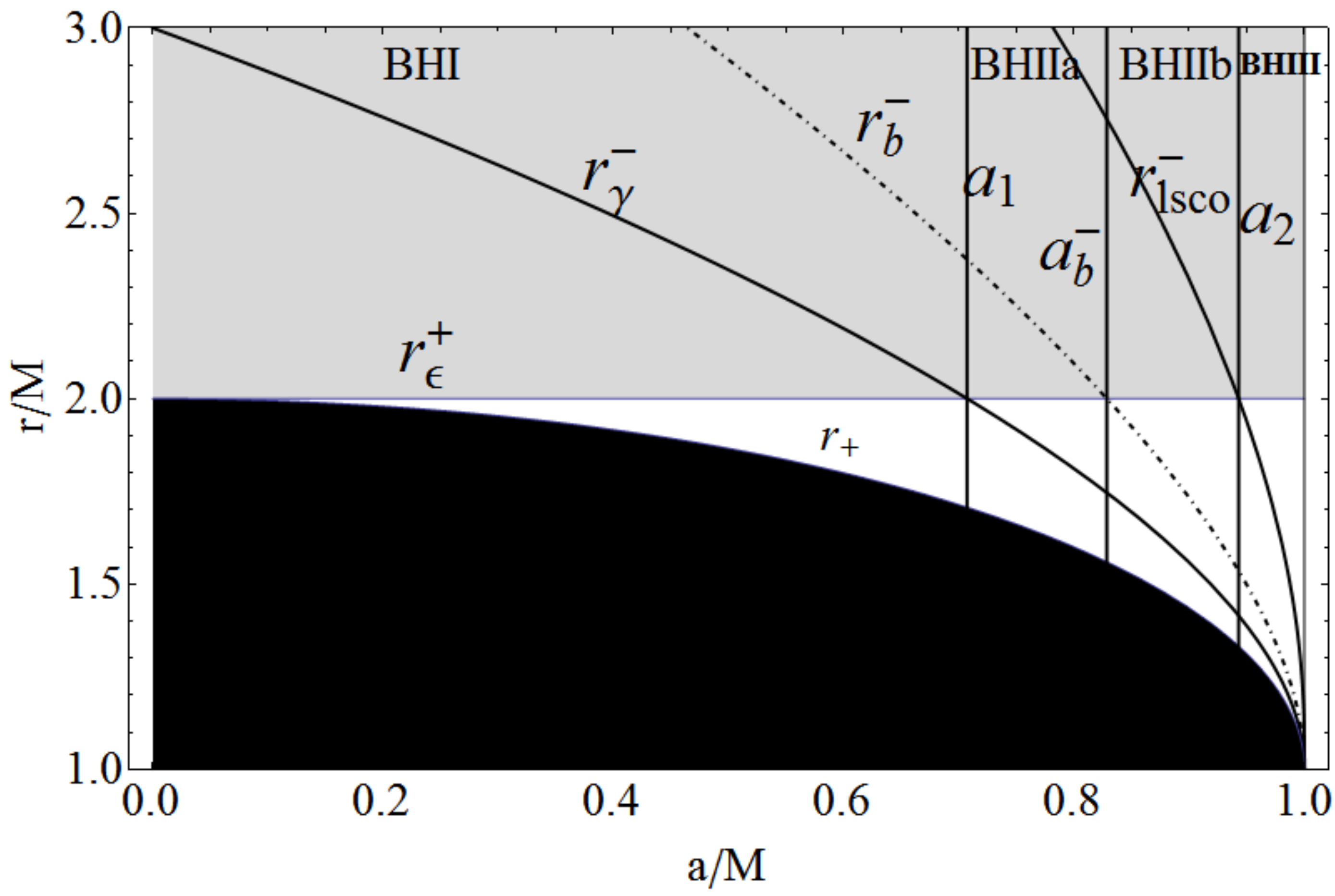}
\\
\includegraphics[scale=0.3]{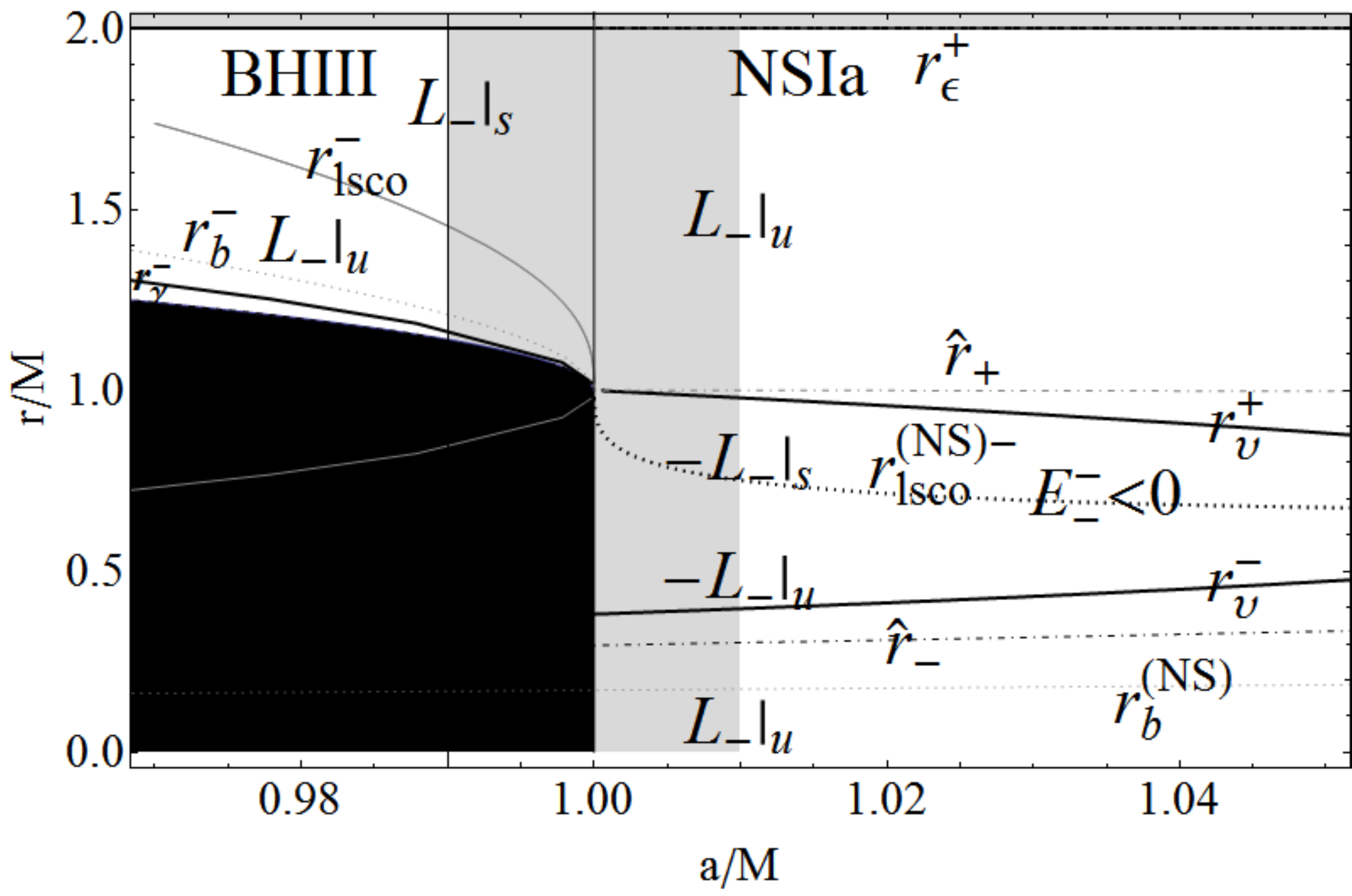}
\end{tabular}
\caption[font={footnotesize,it}]{\footnotesize{Arrangement of the radii $r_i\in\mathcal{R}_{BH}\cup\mathcal{R}_{NS}$ determining the properties
of circular orbits around a rotating central mass and classes of attractors for black hole $(\mathbf{BH})$ (upper and bottom panels) and naked singularity $(\mathbf{NS})$ sources (bottom panel),  as  given in Table\il(\ref{Table:asterisco}).  $(r_b^-,r_b^{(NS)})$ are the marginally bounded orbits, $(r_{lsco}^-,r_{lsco}^{(NS)-})$ are the marginally stable orbits, $r_{\gamma}^-$ is the photon (last circular) orbit. The region outside the static limit $r>r_{\epsilon}^+$ is in gray, and the one  with $r<r_+$ is in black.  The bottom panel shows in light gray  a neighborhood of the geometry $a=M$ between  the \textbf{BHIII} and \textbf{NSIa} classes. The orbital stability in each  region is analyzed and explicitly shown: the  angular momentum is quoted with  its stability property as $\left.\right|_s$  for stable and  $\left.\right|_u$ for unstable orbits.
{Radii $\hat{r}_{\pm}$ are zero angular momentum orbits ($L=0$),  and the  radii $r_{\upsilon}^{\pm}$, zero energy circular orbits,  are counterrotating  ($L=-L_-$) orbits with $E=0$. The region
$]r_{\upsilon}^{-},r_{\upsilon}^{+}[$ contains only counterrotating orbits with $E<0$.}}}
\label{rewpoett}
\end{figure}

The dynamical structure of $\Sigma_{\epsilon}^+$  in $\mathbf{BH}$  spacetimes is determined by the radii  $r_i\in \mathcal{R}_{BH}\in \Sigma_{\epsilon}^+$, as shown in Table\il\ref{Table:asterisco}. The properties of circular orbits outside the ergoregion of black hole geometries are sketched in Fig.\il\ref{rewpoett}, and have been analyzed in detail in \cite{Pu:Kerr}. In the ergoregion, \textbf{BH}-sources are characterized by a unique family of corotating orbits with $ L = L_-$.
We  summarize the results obtained from the analysis of $\Sigma$ by discussing the  three classes of $\mathbf{BH}$ geometries:
$ \mathbf{BHI}: a\in[0,a_1]$, $\mathbf{BHII}=\mathbf{BHIIa}\cup\mathbf{BHIIb}: a\in] a _1 ,a_2]$, where the subclasses  $\mathbf{BHIIa}\equiv] a _1 ,a_{b}^-]$  and $\mathbf{BHIIb}\equiv ]a_{b}^-,a_2]$ and  finally $ \mathbf{BHIII}: a\in]a_2,M]$ (see also Table\il\ref{Table:BHSIGMA1}).
\begin{table*}
\centering
\caption{\footnotesize{Dynamical structure of the ergoregion $\Sigma_{\epsilon}^+$  and length $\mu$ of its  sections in the black hole geometries. The label $()_\varnothing$ refers to the region where circular orbits are not allowed, $()_s$ refers to the  regions where stable circular orbits are possible, $()^<_u$ is for regions where the orbits are unstable with energy  $E<1$ (in units of the mass particle), and finally $()^{\geq}_u$ is for the regions where unstable circular orbits are with $E>1$. The corresponding length is denoted by $\mu_{()}$. The extremes of the ranges  for  the  lengths $\mu_{()}$ are evaluated on the  extreme  geometries of each   spacetime class. The regions are ordered according to the decreasing orbital distance from the source, that is, from the  more distant from the  source to the  closest. The lengths are in units of mass $M$. See also Fig.\il\ref{rewpoett}.
}}\label{Table:BHSIGMA1}
\begin{tabular}{lrcl}
 \hline
\textbf{Black hole geometries:}
\\
 \hline 
\textbf{BHI}:
\(\Sigma_{\epsilon}^+=\Sigma_{\varnothing}\)
\\
\textbf{Length of the sections:} $\mu_{\epsilon}^+=2M$
\\
 \hline 
\textbf{BHIIa}:
\(\Sigma_{\epsilon}^+=\Sigma^{\geq}_u(L_-)\cup\Sigma_{\varnothing}\)
\\
\textbf{Length of the sections:} $\mu_u^{\geq}\in[0,0.253841]$
\\
\hline 
\textbf{BHIIb}:
\(\Sigma_{\epsilon}^+=\Sigma^{<}_u(L_-)\cup\Sigma^{\geq}_u(L_-)\cup\Sigma_{\varnothing}\)
\\
\textbf{Length of the sections:}
$\mu_u^{<}\in[0,0.464516]$,  $\mu_u^{\geq}\in[0.253841,0.120867]$, $\mu_{\varnothing}\in[0.771974,0.417721]$
\\ \hline 
\textbf{BHIII}:
\(\Sigma_{\epsilon}^+=\Sigma_{s}(L_-)\cup\Sigma_{u}^{\geq}(L_-)\cup\Sigma_{u}^{<}(L_-)\cup
\Sigma_{\varnothing}\)
\\
\textbf{Length of the sections:}
$\mu_{s}\in[0,1]$,
$\mu_u^{<}\in[0.464516,0]$,
$\mu_u^{\geq}\in[0.120867,0]$,
$\mu_{\varnothing}\in[0.417721,0]$
\\ \hline
\end{tabular}
\end{table*}
\begin{figure}%
\begin{tabular}{cc}
\includegraphics[scale=.3]{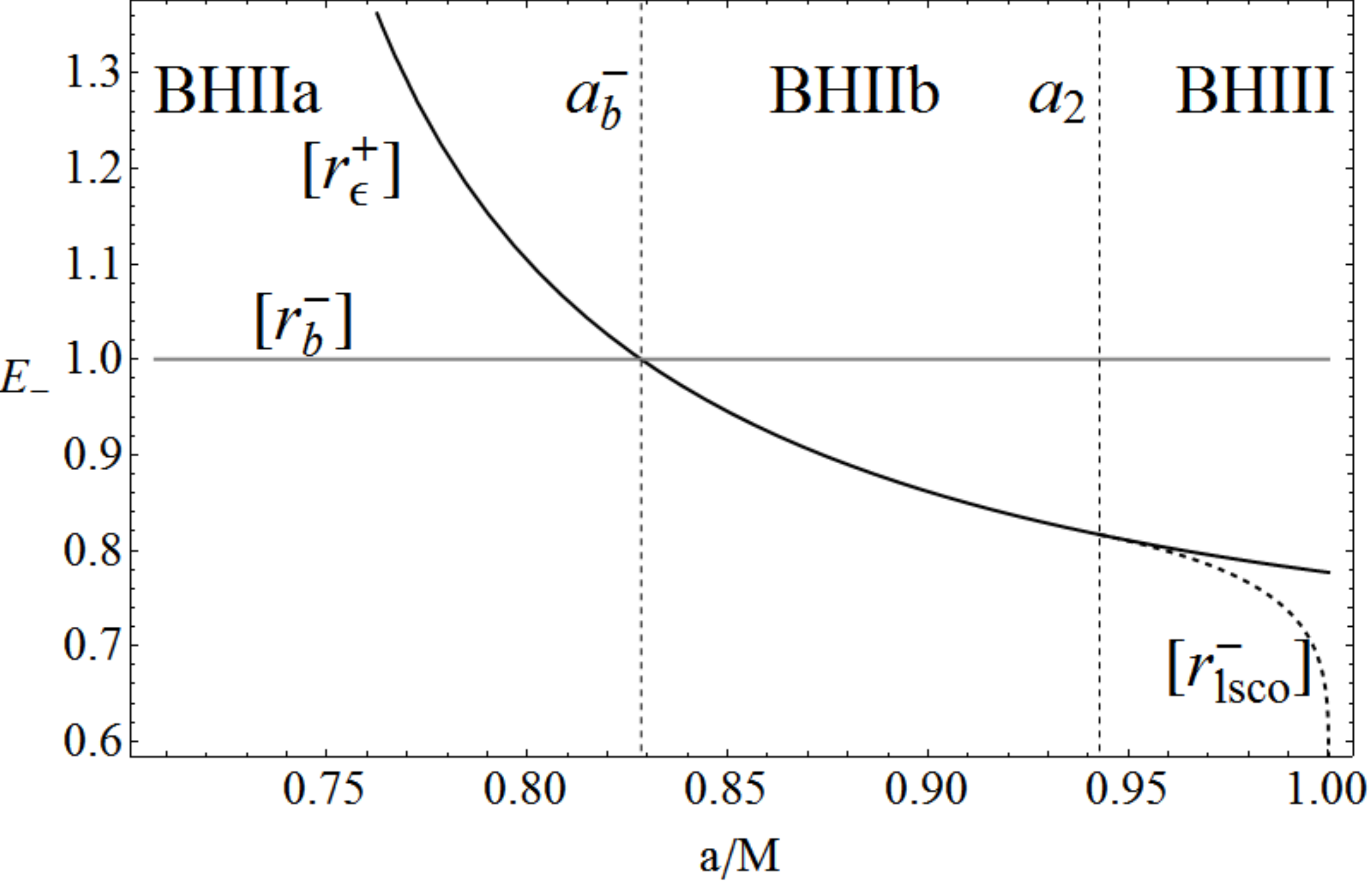}
\\
\includegraphics[scale=.3]{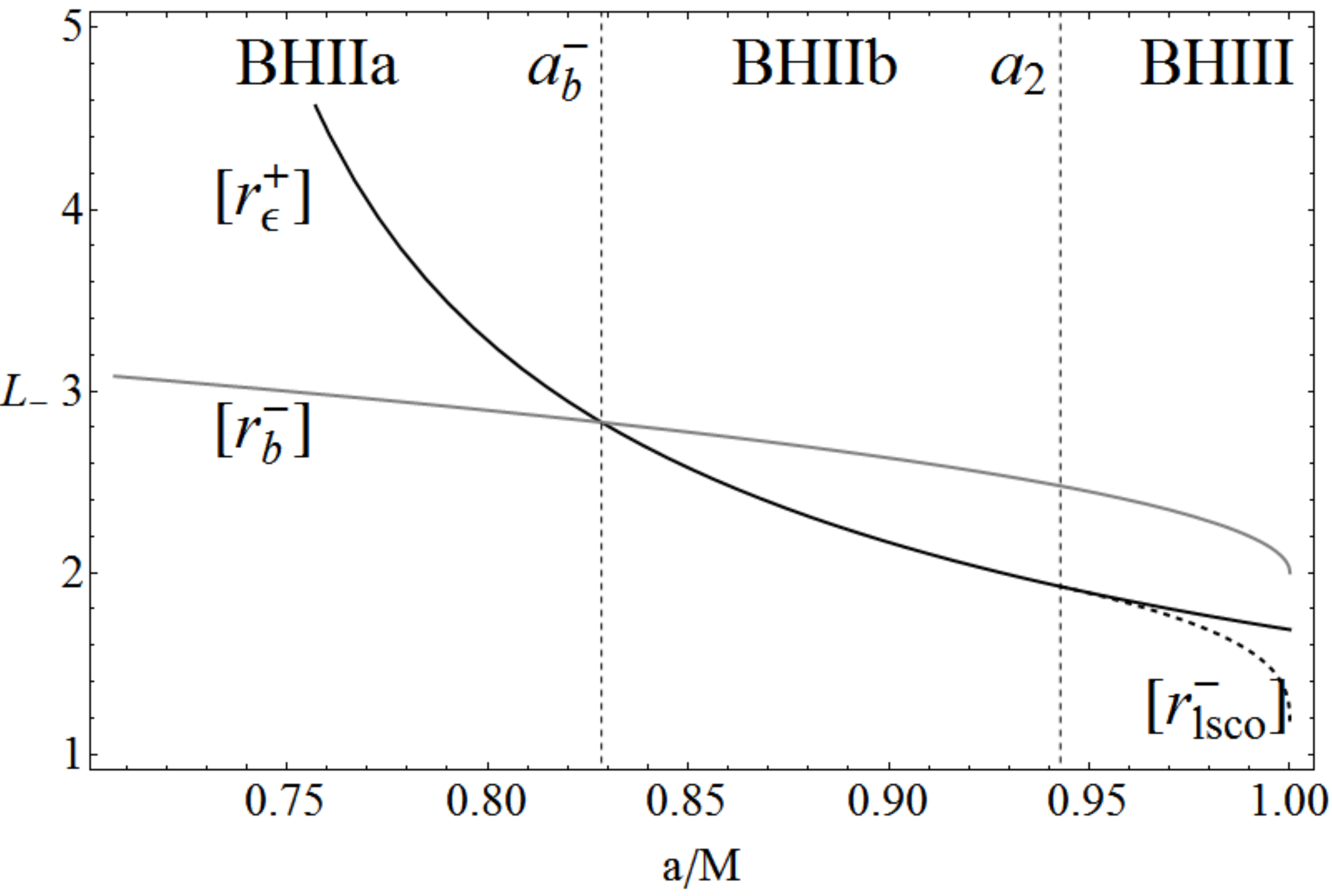}
\end{tabular}
\caption[font={footnotesize,it}]{\footnotesize{Black hole case:
energies $E_-$ (upper panel)  and   the angular momentum  $L_-$ (bottom panel) as functions of the spin-mass ratio $a/M$ of the black hole. The energies are in units of particle mass $\mu$ and the angular momentum in units of $M\mu$. The spins $a_b^-= 0.828427M$, $a_2\equiv {2 \sqrt{2}}/{3}M$ are denoted by dashed lines.  The curves are the energies and the angular momentum respectively on specific orbits represented  in square brackets next to the curves.
$r_{\epsilon}^+$ is the static limit (black curve), $r_b^-$ is the marginally bounded orbit (gray curve), $r_{lsco}^-$ is the last stable circular orbit (dashed curve).
 }}
\label{Fig:PlotVBHRic}
\end{figure}
\subsubsection{The class $\mathbf{BHI}: a\in[0,a_1]$}
The class   $\mathbf{BHI}$  is bounded by  the Schwarzschild static spacetime  and the spacetime with spin $a=a_1\approx0.707107M$.
No circular motion  can occur for $r<r_{\epsilon}^+$, it is then $\Sigma_{\epsilon}^+=\Sigma_{\varnothing}$:  the orbital amplitude of this region reaches its maximum  in the limit geometry ${a=a_1}$, the static limit in this spacetime coincides with the photon orbit
(see Fig.\il\ref{rewpoett} and the discussion in Sec.\il\ref{sec:stl}).
There is no solution for Eq.\il(\ref{Eq:Kerrorbit}),
 because the  effective potential
 is always increasing $(V'>0)$ in this region.
A particle penetrating the ergoregion with $\dot\theta=0$ and $\dot r<0$ must move along the radial direction with a twisting along $\phi$.
There is  a belt,  with  boundary at $r_{\gamma}^-$,   surrounding the static limit (in $r>r_{\epsilon}^+$) where no  inner edge for a thin disk (P-W point of accretion) and no  cross point for the origin  of  matter Jets can exist in the (outer) neighboring regions of the static  limit. This forbidden region, however, decreases  in extension as the spin-mass ratio approaches  the upper limit of this class; its maximum value is  in the limiting  static geometry where its length is  equal to the mass  $M$ of the attractor. For  spin values close to  $a=a_1$  the forbidden belt decreases in length, until  the launching point of the matter Jet  approaches the static limit at $r\gtrapprox r_{\epsilon}^+$.
\subsubsection{The class $ \mathbf{BHII}: a\in]{a_1},{a_2}]$}
The dynamical structure of the ergoregion in the case of $\mathbf{BHII}$ geometries is
shown in Table\il\ref{Table:BHSIGMA1}.
The instability region contains  the orbit $r_b^-$;  it is therefore convenient to  consider separately the  sub-class  $\mathbf{BHIIa}: a\in]{a_1},{a_b^-}]$  and $\mathbf{BHIIb}: a\in]{a_b^-},a_2] $.  For a spacetime  with $a_b^-= 0.828427M$ it is $r_{b}^-=r_{\epsilon}^+$.

For $\mathbf{BHIIa}$ geometries with  $a\neq a_b^-$ it is $ \Sigma_u(L_-)= \Sigma^{\geq}_u(L_-)$ which is a region with unstable unbounded $(E_->1)$ orbits whose energy increases as the orbit approaches the photon orbit at $r_{\gamma}^-$. These orbits are unstable and under  a perturbation they can eventually escape with $\dot{r}>0$ and $\dot \phi>0$, crossing the ergoregion at a certain finite time {$\bar{\tau}:\;{r}(\bar{\tau})=r_{\epsilon}^+$ and  $\dot{r}(\bar{\tau})>0$  and $\dot{\phi}(\bar{\tau})>0$}.
The particle might run away to infinity, because the energy to decay into a lower circular orbit is higher than the  energy required for spiraling  outward to an exterior orbit. This could give rise to the ejection of positive energy particles outside the static limit. If we consider a thick accretion disk model, we could say that in the regions adjacent to the static limit or $]r_{\epsilon}^+,r_b^-]$  there can be only open configurations with a cross point in that region (or even on the static limit; see section  Sec.\il\ref{sec:stl}), where there  is a  (critical) minimum pressure value, with the consequent launch of funnels of matter along the axis of symmetry.
In fact, the cross point of an open configuration, i.e. the origin point  of (corotating) matter Jets,  can occur in the ergoregion of $\mathbf{BHIIa}$ geometries  in   $\Sigma^{\geq}_u(L_-)\in\Sigma_{\epsilon}^+$,  where   the matter, twisting with the source (i.e.  with initial angular momentum  $L=L_-$),  will be aligned with the rotation axis, while a part of the  trapped particles in the ergoregion  can be expelled  outside, without falling into the singularity. {As shown in } Table\il\ref{Table:BHSIGMA1}, the orbital extension  $\mu^{\geq}_u$ is less than  $\mu^{\geq}_u=0.26 M$ and the horizon is then covered by the orbit of the photon at $r_{\gamma}^-$.
However, the static limit crosses  smaller region of instability  as the attractor spin  approaches the boundary of the class  \textbf{BHIIa},
  where close to the static limit, as  $a=a_{b}^-$,  there are  accretion   points  of  closed surfaces according to the P-W accretion mechanism.

\begin{figure}
\begin{tabular}{cc}
\includegraphics[scale=.3]{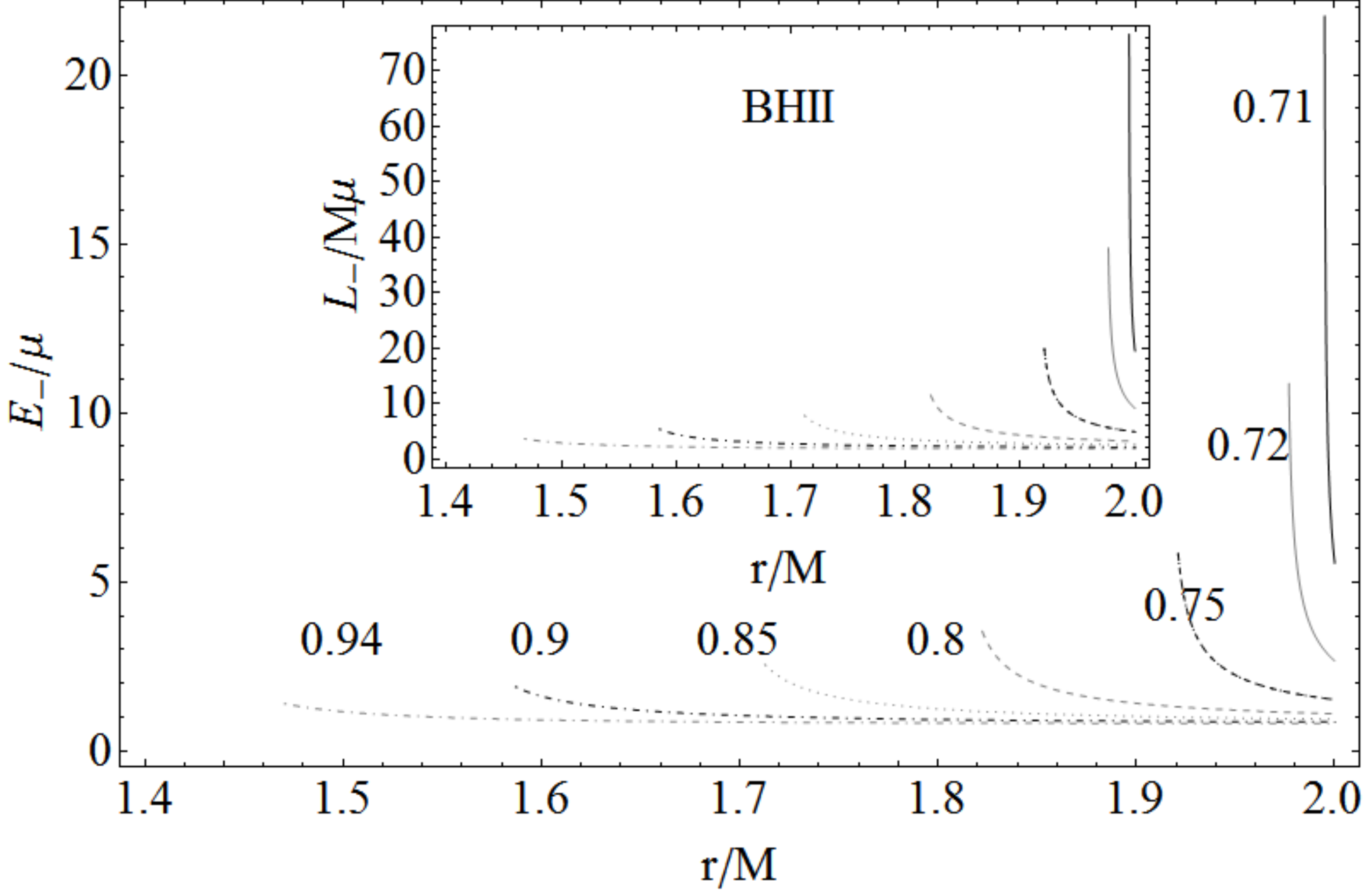}
\end{tabular}
\caption[font={footnotesize,it}]{The energy $E_-$  and the corresponding  angular momentum $L_-$ (inside panel) as functions of  $r/M$ and for different  values of the
spin-to-mass ratio of {\textbf{BHII}} sources.
}
\label{Fig:red-cra}
\end{figure}

 In $\mathbf{BHIIb}$ spacetimes,
the region $\Sigma_{u}^{<}\subset \Sigma_{\epsilon}^+$  locates
the instability points of the toroidal  thick  configuration of  corotating matter   in accretion. The hydrostatic pressure on the cusps is minimum, and  these points  can be even on the static limit (see Sec.\il\ref{sec:stl}). No toroidal equilibrium disk    can be  inside the ergoregion  due to the gravitational and hydrostatic instability in $\Sigma_u^{<}$.
The point of accretion can be also very close to the static limit. Only open funnels are possible from a  point in $\Sigma_{u}^\geq$, where  the launch of a Jet is possible.  The length of   $\Sigma_{u}^\geq$  decreases, as the rotation of the attractor  increases, approaching  the horizon and  creating therefore a gorge whose  extension decreases from the maximum of  $\mu_{u}^\geq\approx 0.12M$ (see Table\il\ref{Table:BHSIGMA1}).
In the proximity of the static limit  ($r\gtrapprox r_{\epsilon}^+$) in \textbf{BHIIb} spacetimes,  closed configurations in equilibrium are possible  as $a\lessapprox a_2$.
The situation becomes critical as the attractor spin increases up to the upper limit of the \textbf{BHIIb} class with $a=a_2$. Then, the  closed configurations in equilibrium  approach their center of maximum hydrostatic pressure, until the center coincides with the static limit in the spacetime with $a=a_2$. Then
 the inner edge of the disk would be  within the region $\Sigma_u^{\geq}$.
The energy $E_- > 0$  and  orbital angular momentum $L_->0$ increase, approaching the limiting value at $r_{\gamma}^-$ (see also Fig.\il\ref{Fig:red-cra}). On the other hand,   at a fixed orbit $r$,  the energy decreases   as the attractor spin  increases. As shown in  Fig.\il\ref{Fig:PlotVBHRic}, the minimum energy extractable during an orbital decay in $\mathbf{BHIIa}$ geometries is  $E=\mu$ on the orbit $r_b^-$. The maximum extractable energy from an initial decay by a particle in  $\Sigma_{u}^{\geq}$ does not exceed the value $E=1.6\mu$  corresponding to a particle that decays from an orbit very close to the static limit.
In $\textbf{BHIIb}$ spacetimes, the minimum extractable energy is not less than $E=0.8\mu$ as shown in Figs.\il\ref{Fig:PlotVBHRic}.

\subsubsection{The class $ \mathbf{BHIII}: a\in]a_2,M]$}
In $ \mathbf{BHIII}$   spacetimes, there can be stable orbits in the ergoregion $\Sigma_{\epsilon}^+$
(see Fig.\il\ref{rewpoett} and Table\il\ref{Table:BHSIGMA1}).
As it is  $r_{\epsilon}^+\in\Sigma_{s}$, then the static limit can correspond to  a stable orbit, according to the discussion of Sec.\il\ref{Sec:pro}
(see also Sec.\il
\ref{sec:stl}). It can exist a stable thin ring of dust orbiting   very close around   $r_{\epsilon}^+$ ($r=r_{\epsilon}^+\pm\epsilon$ with $\epsilon\gtrapprox0$). The inner edge and the center (maximum of the hydrostatic pressure) of the thin disk can be in $\Sigma_s\subset\Sigma_{\epsilon}^+$.
The length  $ \mu_s  $ of this region increases  as  $a\lessapprox M$.
A thick corotating accretion disk in $\mathbf{BHIII}$ spacetimes can be entirely  inside  $\Sigma_{\epsilon}^+$, dynamically including the  ring formation  in equilibrium up to the  accretion and the  Jet  launch.
 Thus the point of launch of the  Jet, according with P-W mechanism, can be located  in a region very close to the horizon, as the attractor spin increases. A slightly change in the attractor  spin would produce a transition from a section to another inside the ergoregion, causing  a change in the stability properties   of the orbital matter.
However,  the maximum  length  $\mu_s$ of the region of stability is $\mu_s=M$; therefore, the elongation of a thin disk  entirely contained in $\Sigma_{\epsilon}^+$ cannot be greater than $M$.

As the largest section for a fixed attractor  is $ \Sigma_{u}^<$, one could say that closed stable configurations are favored in  \textbf{BHIII} spacetimes, Jets
in $\textbf{BHIIa}$ sources, and the accretion disks  in  $\mathbf{BHIIb}$ and  $\mathbf{BHIII}$ attractors.
The behavior of the energy and angular momentum in these spacetimes is shown in  Figs.\il\ref{Fig:PlotVBHRic}.  The maximum energy can be extracted from a particle that decays from  an orbit close to the static limit and slightly greater than $E\approx 0.8\mu$, while the angular momentum does not exceed the value $L\approx 2 M\mu$.
\subsection{Naked singularities}
\label{sec:nss}
We distinguish three classes of naked singularity sources:
$ \textbf{NSI}: a\in]1,{a_3}]$,   split into \textbf{NSIa}: $a\in]1,{a_\mu}]  $  and \textbf{NSIb}: $a\in]{a_\mu},{a_3}]$, the class
$\mathbf{NSII}: a\in]{a_3},{a_4}]$, and finally the class
$\mathbf{NSIII}: a\in]{a_4},\infty]$,   split into  $\mathbf{NSIIIa}: a\in]{a_4},a_b^{NS}]$ and $\mathbf{NSIIIb}: a\in]a_b^{NS}, \infty[$ (see also
Table \ref{Table:asterisco}). A characterization of the orbital  energy in $r>r_{\epsilon}^+$ can be found with some details in \cite{Pu:Kerr}.
In contrast with \textbf{BH}, in \textbf{NS}  spacetimes the singularity  is always covered by the region  $\Sigma_{u}^{\geq}(L_-)$, with  unstable and unbounded circular orbits.
\begin{table*}
\centering
\caption{\footnotesize{Dynamical structure of the ergoregion $\Sigma_{\epsilon}^+$  and length $\mu$ of its  sections in naked singularity  geometries: the label $()_\varnothing$ refers to the regions where circular orbits are not allowed,  $()_s$ to regions with stable circular orbits, $()^<_u$ to regions with  unstable orbits with energy  $|E|<1$ (in units of the mass particle), and finally $()^{\geq}_u$ is for the regions where unstable circular orbits exist with $|E|>1$.
The corresponding lengths are denoted by $\mu_{()}$.  The boundaries of the ranges  for   the length $\mu_{()}$ correspond to values of the boundary  geometries of each spacetime class. The regions ${\Large{\not{\Sigma}}}$, and the corresponding length ${\Large{\not}}{\mu}$, refer to orbits with negative energy.
The regions are ordered according to the decreasing orbital distance from the source, that is, from the  more distant from the  source to the  closest. The lengths are in units of mass $M$. For details see
 Figs.\il\ref{rewpoettx}.
The  angular momentum characterizing the possible orbits are in round brackets. There are not regions with both   corotating  $(L_-)$  and counterrotating  $(-L_-)$ angular momenta. The two kinds of orbits  are  confined in disjoint regions.}}
\label{Table:NS}
\begin{tabular}{lrcl}
 \hline
\textbf{Naked singularity geometries:}
\\
 \hline 
\\
\textbf{NSIa:}
\(\Sigma_{\epsilon}^+=\Sigma_{u}^-(L_-)\cup\Sigma_s(-L_-)\cup{\Large{\not}\Sigma_{s}}(-L_-)
      \cup{\Large{\not}\Sigma_{u}^<}(-L_-)\cup\Sigma_{u}^<(-L_-)
      \cup\Sigma_{u}^<(L_-)\cup\Sigma_{u}^{\geq}(L_-)\)
\\
\textbf{Length of the sections:}
$\mu_{u}^<(L_-)\in[1,1.00873]$,
$\mu_s(-L_-)\in[0,0.324601]$,
${{{\Large{\not}}{\mu}}}_s(-L_-)\in[0,0]$,
\\
${{{\Large{\not}}{\mu}}}_u(-L_-)\in[0.618034,0]$,
$\mu_u(-L_-)\in[0.0863683,0.298875]$,
$\mu_u^{<}(L_-)\in[0.124025,0.16957]$,\\
$\mu_u^{\geq}(L_-)\in[0.171573,0.198221]$.
\\
\hline
\textbf{NSIb:}
\(
\Sigma_{\epsilon}^+=\Sigma_{u}^-(L_-)\cup\Sigma_u^<(-L_-)\cup\Sigma_s(-L_-)
\cup\Sigma_{u}^<(L_-)\cup\Sigma_{u}^{\geq}(L_-)\)
\\
\textbf{Length of the sections:}
$\mu^{<}_{u}(L_-)\in[1.00873,1.25]$,
$\mu_s(-L_-)\in[0.324601,0]$,\\
$\mu_u(-L_-)\in[0.298875,0]$,
$\mu^{<}_u(L_-)\in[0.16957,0.483478]$,
$\mu^{\geq}_u(L_-)\in[0.198221,0.266522]$
\\
\hline
\textbf{NSII:}
\(
\Sigma_{\epsilon}^+=\Sigma_{s}(L_-)\cup\Sigma_{u}^<(L_-)\cup\Sigma_{u}^{\geq}(L_-)
\)
\\
\textbf{Length of the sections:}
$\mu_s\in[1.25,0]$,
$\mu_{u}^<\in[0.483478,1.08485]$,
$\mu_{u}^{\geq}\in[0.266522,0.915154]$
\\
\hline
\textbf{NSIIIa:}
\(
\Sigma_{\epsilon}^+=\Sigma_{u}^<(L_-)\cup\Sigma_{u}^{\geq}(L_-)
\)
\\
\textbf{Length of the sections:}
$\mu_{u}^<\in[1.08485,0]$,
$\mu_{u}^{\geq}\in[0.915154,2]$
\\
\hline
\textbf{NSIIIb:}
\(
\Sigma_{\epsilon}^+=\Sigma_{u}^{\geq}(L_-)
\)
\\
\textbf{Length of the sections:}
$\mu_{u}^{\geq}=2$
\\
\hline
\end{tabular}
\end{table*}
 \begin{figure}
\begin{tabular}{c}
\includegraphics[scale=0.3]{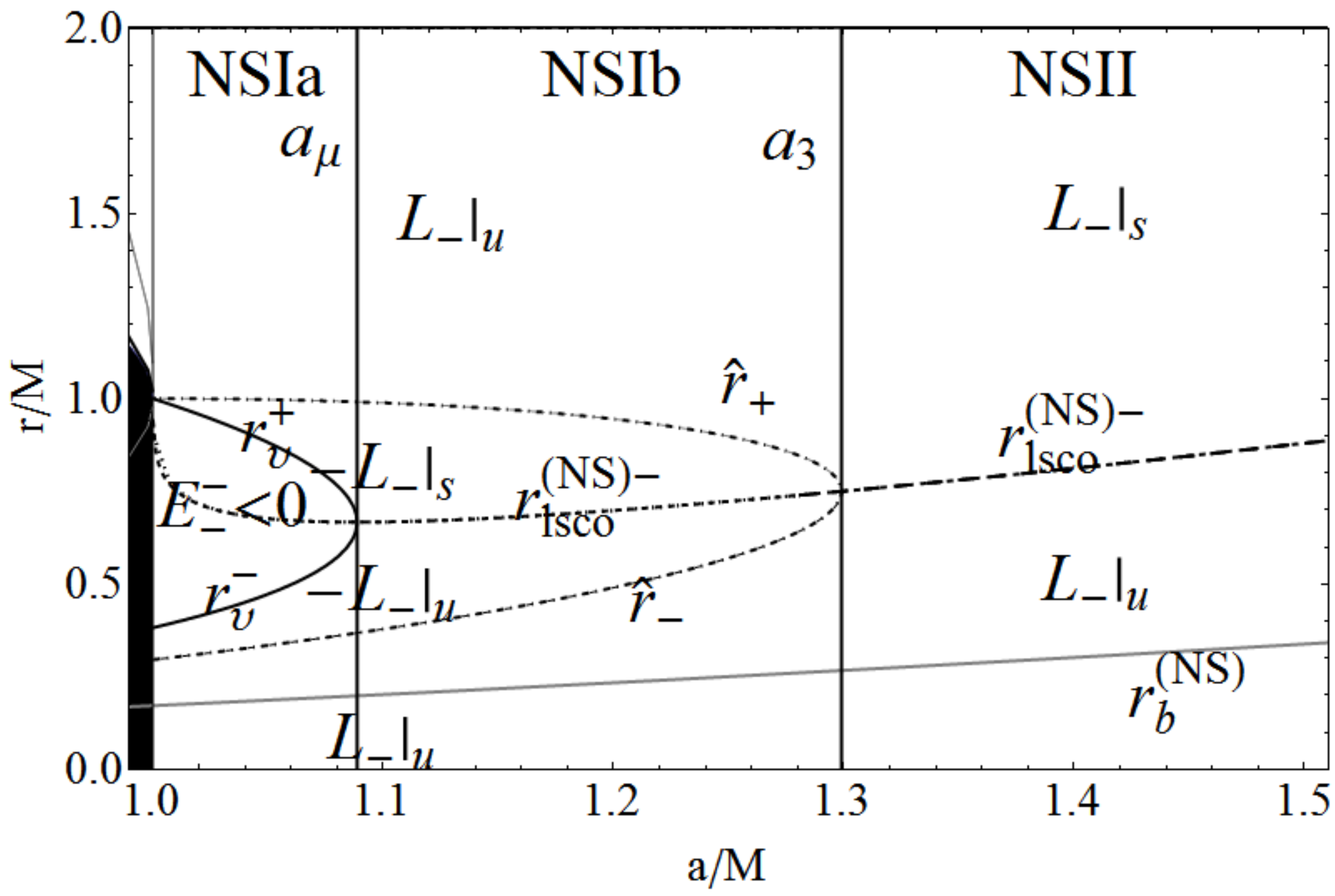}
\\
\includegraphics[scale=0.3]{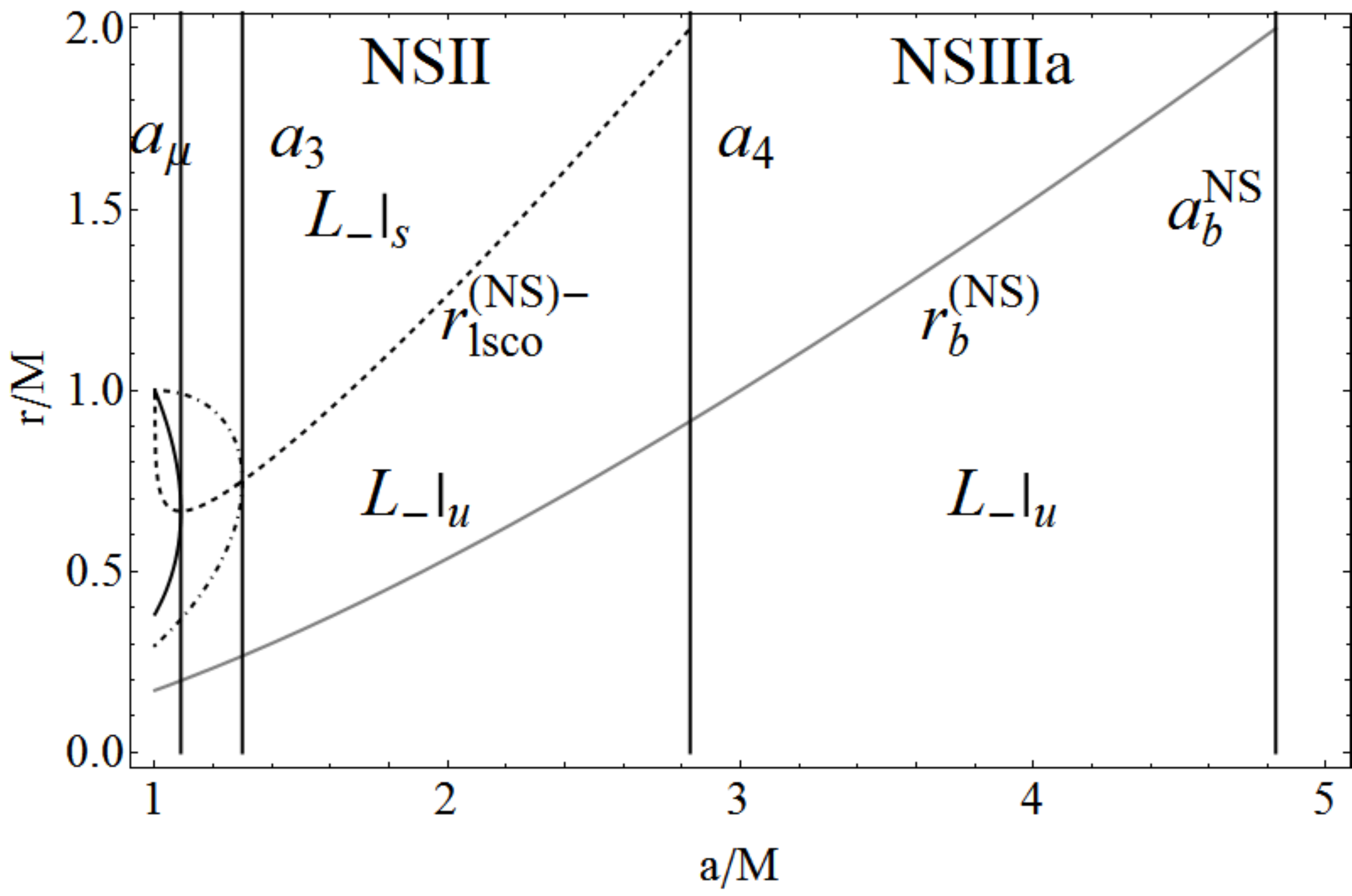}
\end{tabular}
\caption[font={footnotesize,it}]{\footnotesize{Arrangement of the radii  $r_i\in\mathcal{R}_{NS}$ determining the properties
of circular orbits around a rotating  naked singularity $(\mathbf{NS})$ in the region $\Sigma_{\epsilon}^+$.
The orbital stability in each  region is analyzed and explicitly shown. The  angular momentum is quoted with  its stability property as $\left.\right|_s$  for stable and  $\left.\right|_u$ for unstable orbits.  $r_b^{(NS)}$ corresponds to the marginally bounded orbits, $r_{lsco}^{(NS)-}$ to the marginally stable orbits. On
$\hat{r}_{\pm}$, the orbits have zero angular momentum ($L=0$). The  radii $r_{\upsilon}^{\pm}$ correspond to counterrotating ($L=-L_-$)  orbits with zero energy
($E=0$). The region
$]r_{\upsilon}^{-},r_{\upsilon}^{+}[$ contains only counterrotating orbits with $E<0$.}}
\label{rewpoettx}
\end{figure}
\begin{figure}%
\includegraphics[width=0.471\textwidth]{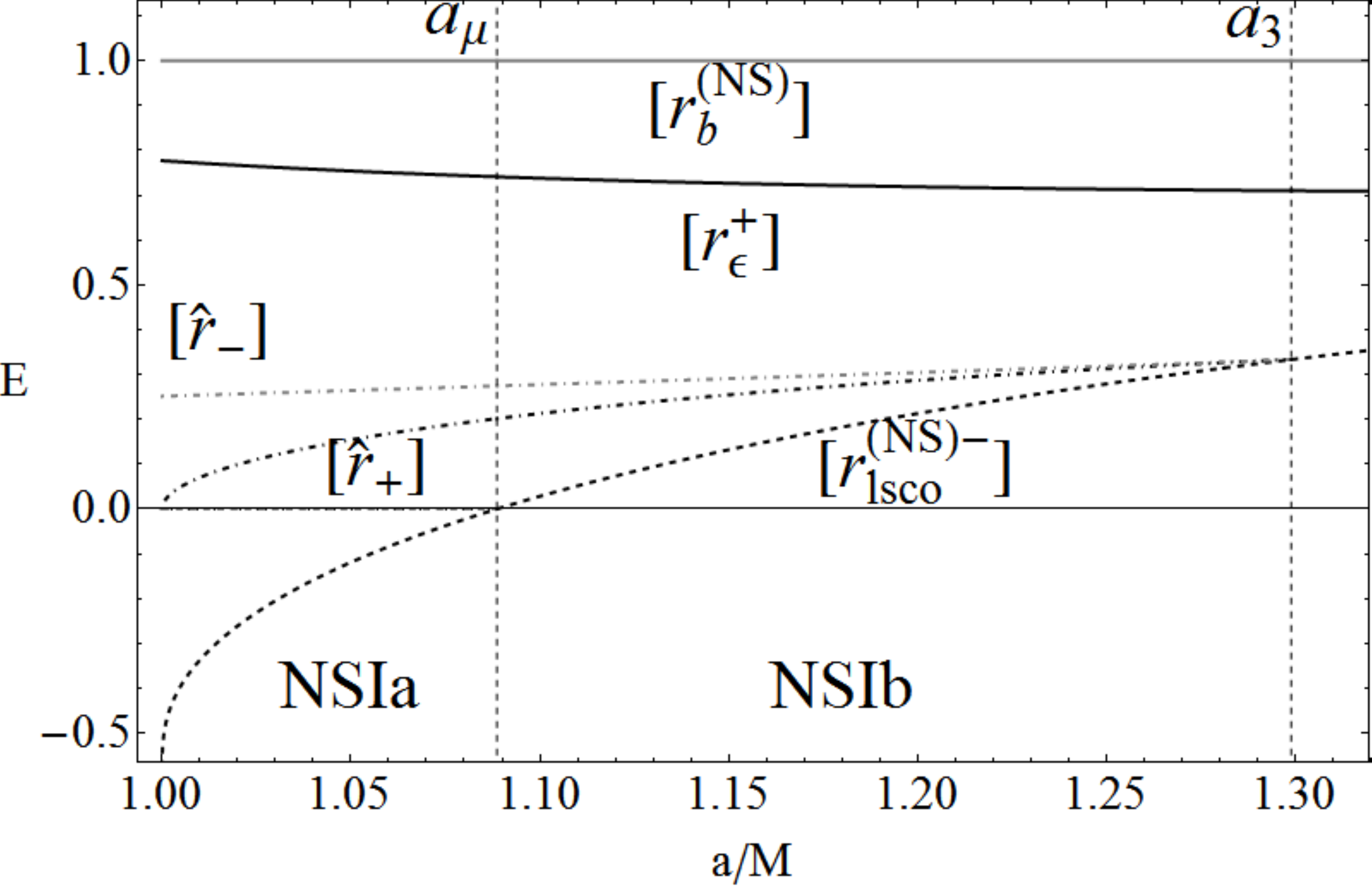}\\
\includegraphics[width=0.471\textwidth]{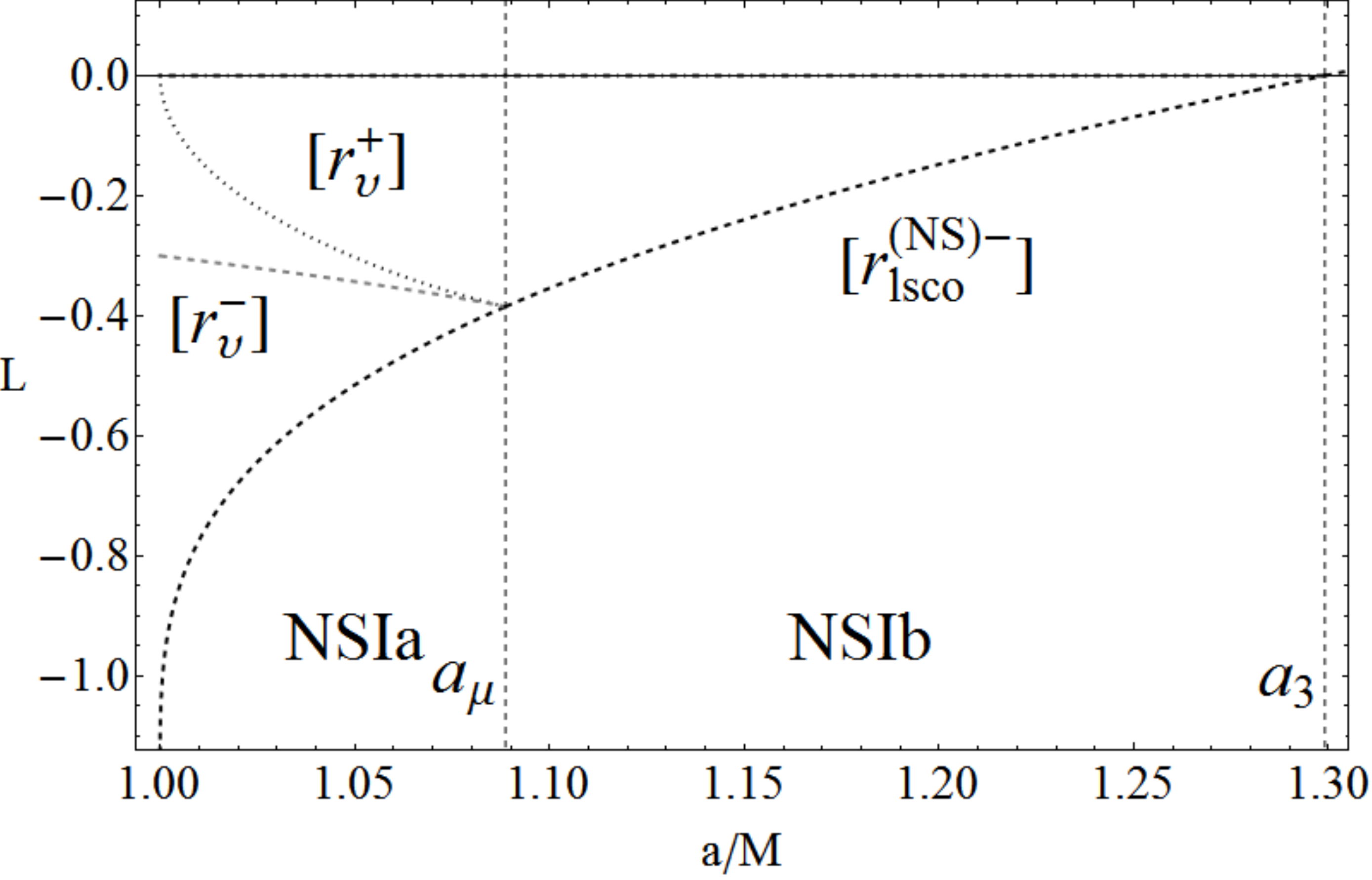}\\
\includegraphics[width=0.471\textwidth]{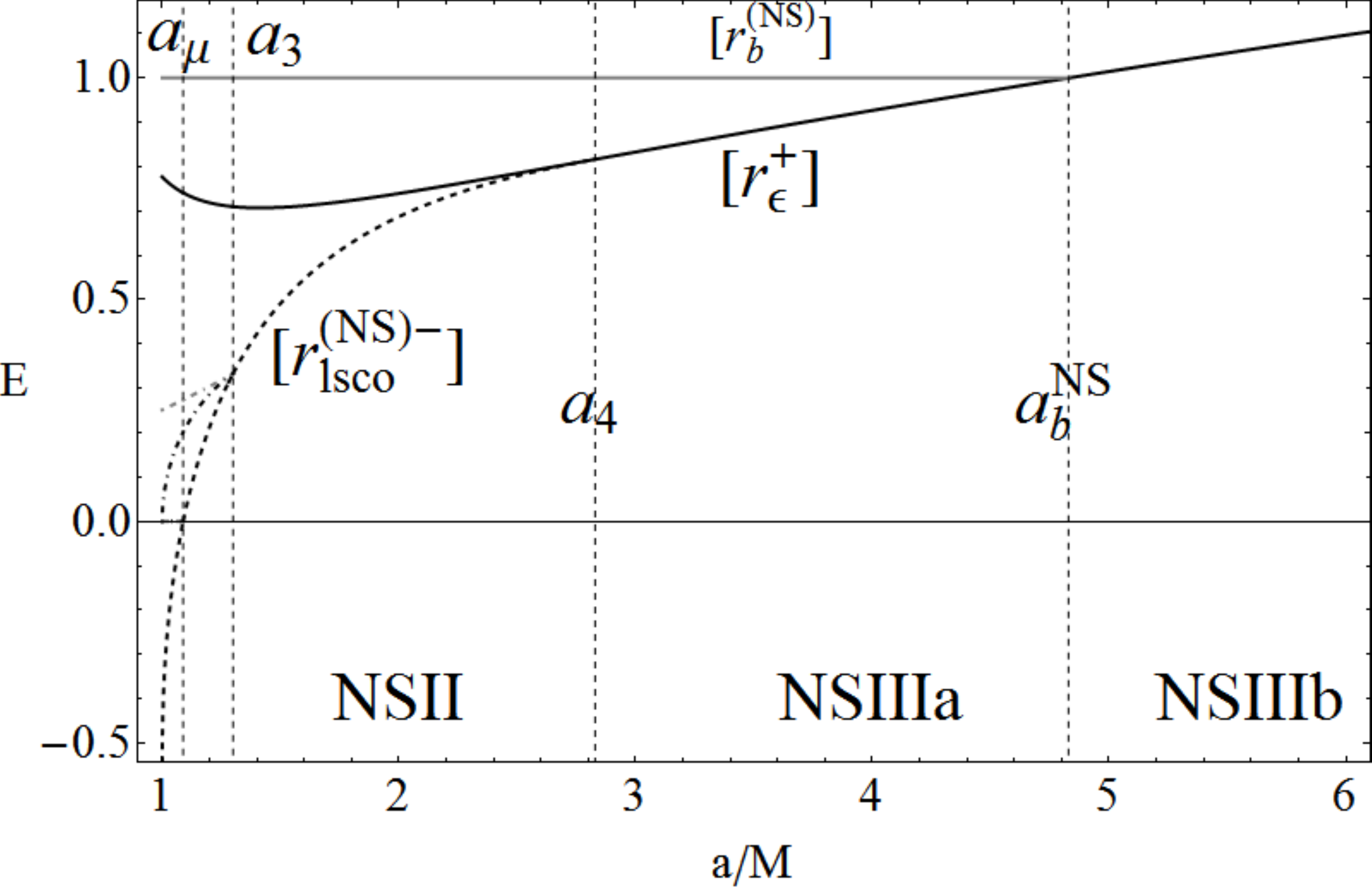}\\
\includegraphics[width=0.471\textwidth]{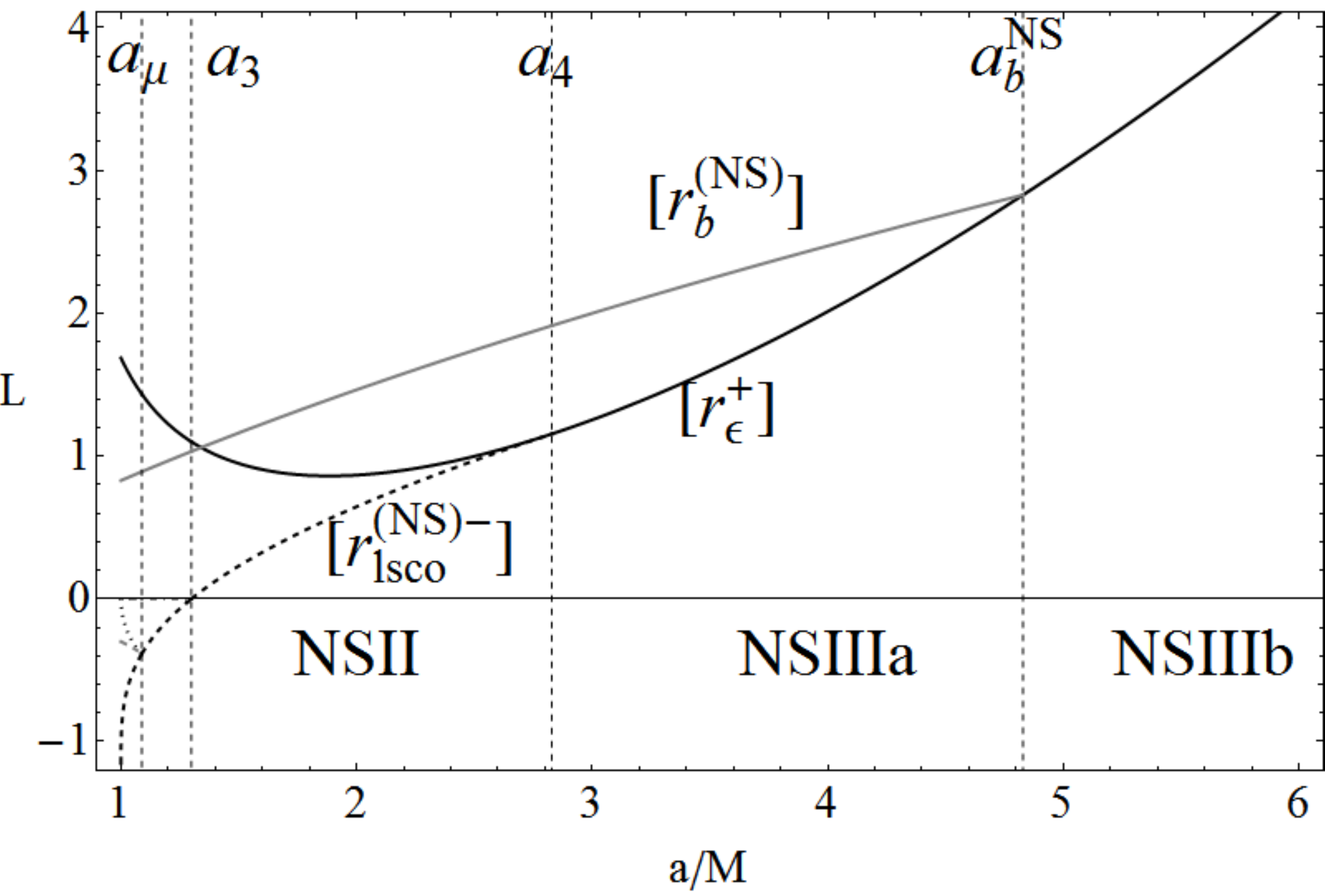}
\caption{\footnotesize{The naked singularity case:
The energies $E$ and  angular momenta  $L$ as functions of the spin-mass ratio $a/M$ of the naked singularity.
The energies and the angular momenta are  in units of the particle mass $\mu$. The spins
${a}_3/M\equiv 3\sqrt{3}/4$, $a_{\mu}/M\equiv {4 \sqrt{{2}/{3}}}/{3} $,
$a_b^{NS} /M  \equiv 2 \left(1+\sqrt{2}\right)\approx4.82843$, ${a_4}/M=2 \sqrt{2}$ are denoted by dashed lines.
The curves represent the energies and the angular momenta, respectively, for specific orbits marked in square brackets next to the curves.
$r_{\epsilon}^+$ is the static limit, $r_b^{(NS)}$ is the marginally bounded orbit, $r_{lsco}^{(NS)-}$ is the last stable circular orbit. On $\hat{r}_{\pm}$, it is $L(\hat{r}_{\pm})=0$, and on $r_{\upsilon}^{\pm}$ (effective ergosurfaces), it is $E(r_{\upsilon}^{\pm})=0$.}}
\label{Exremenagr}
\end{figure}
\subsubsection{The class $ \textbf{NSI}: a\in]1,{a_3}]$}\label{Sec:firstNS}
\begin{figure}
\includegraphics[width=0.5\textwidth]{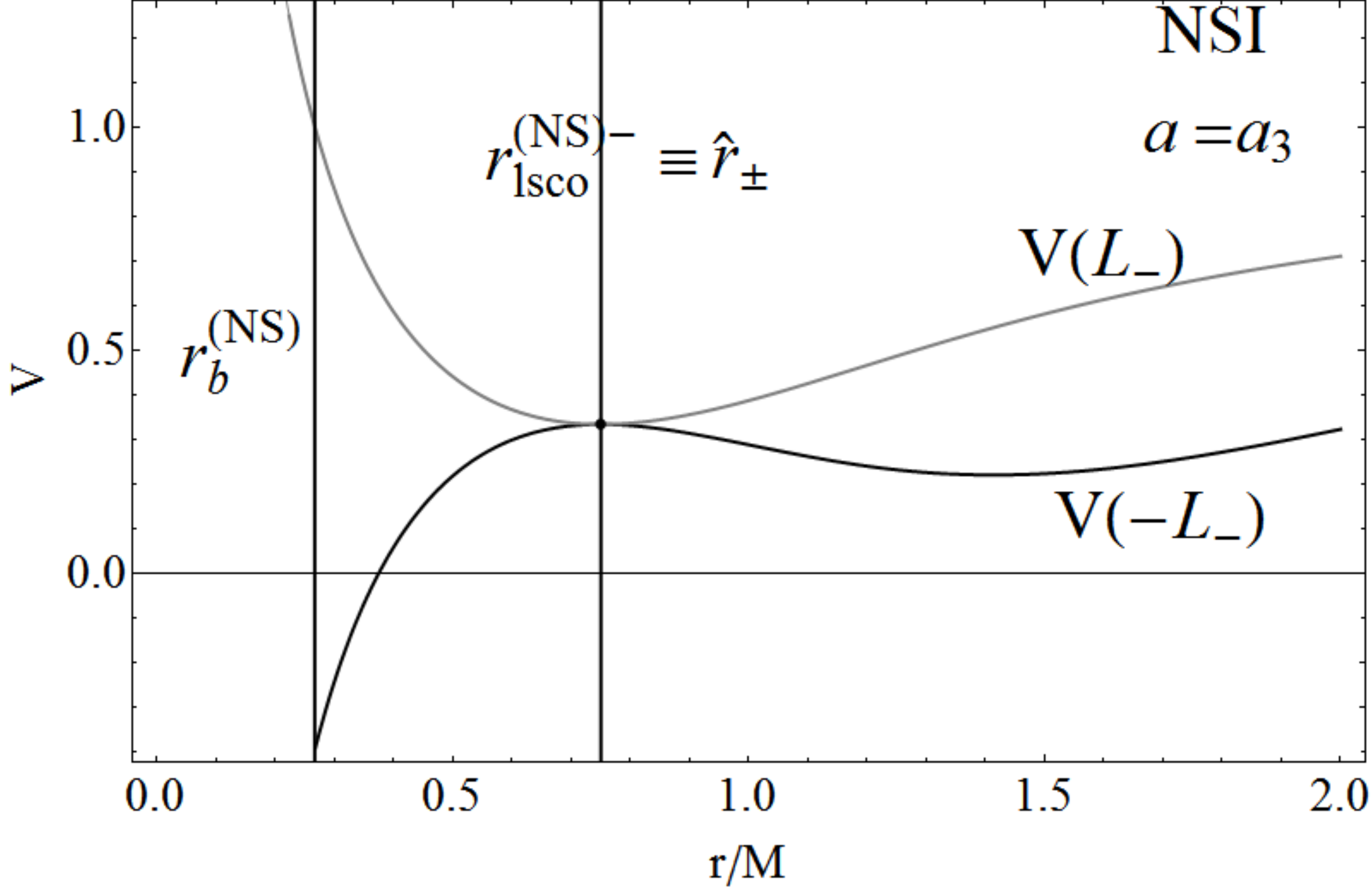}
\\
\includegraphics[width=0.5\textwidth]{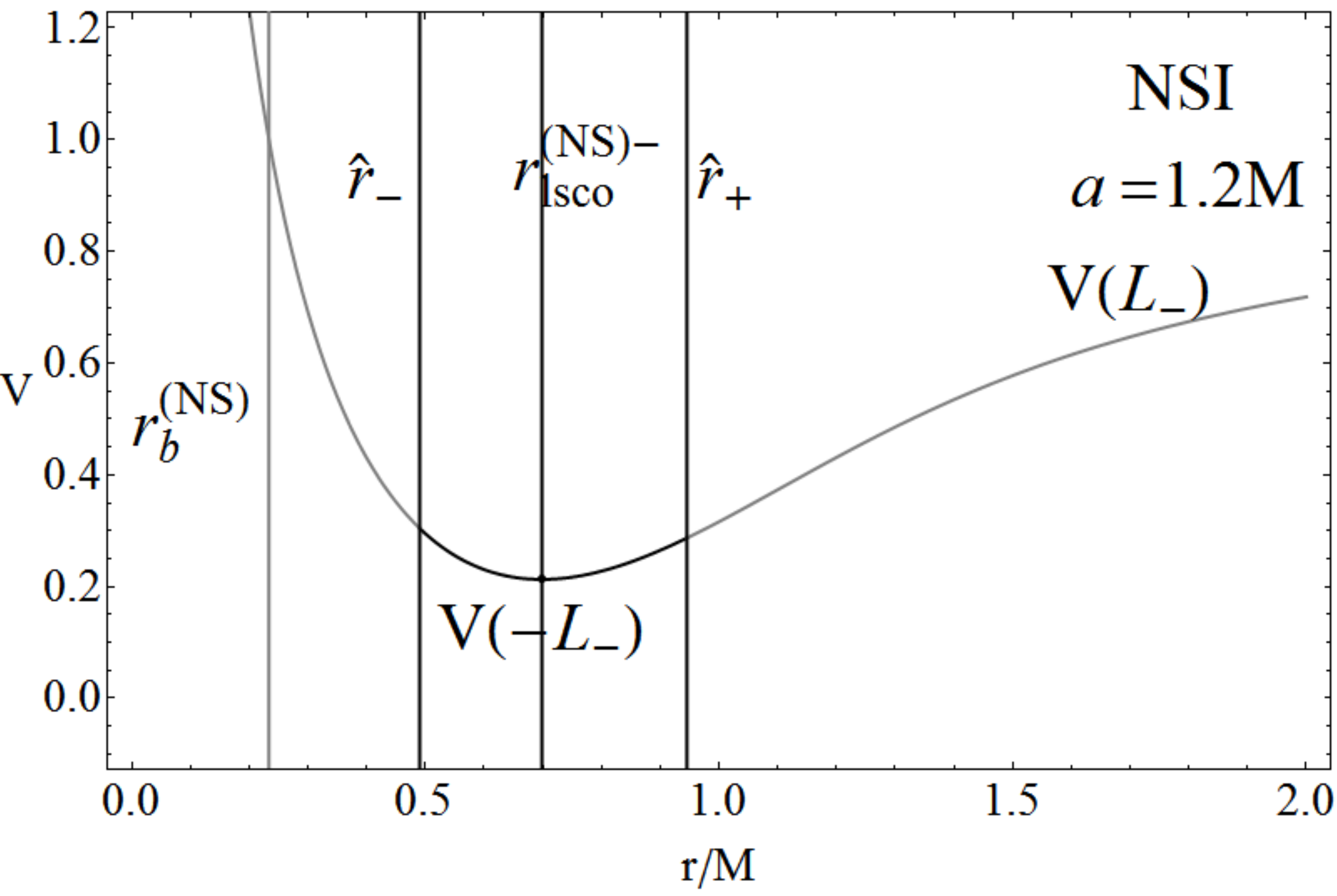}
\\
\includegraphics[scale=.3]{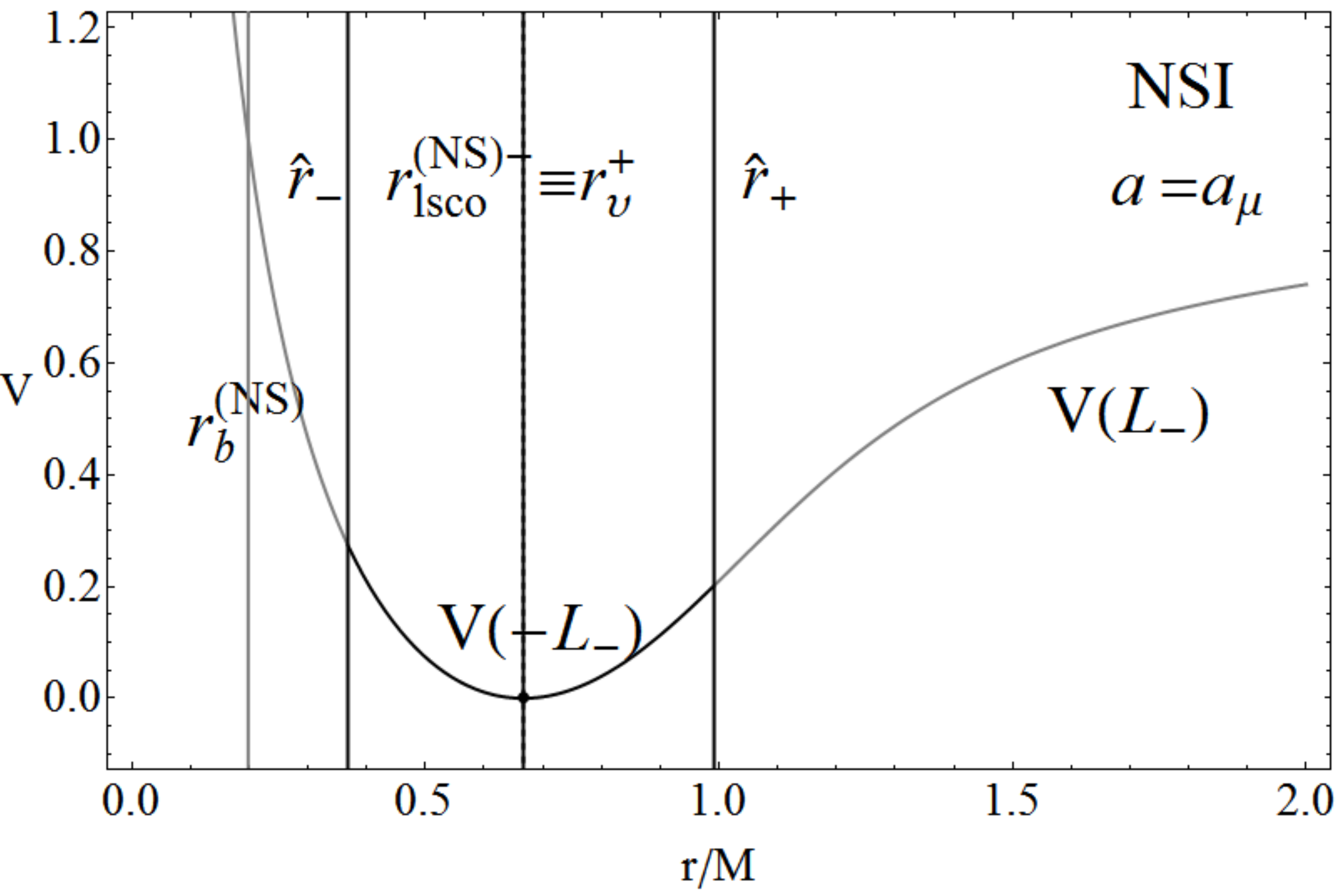}
\\
\includegraphics[scale=.3]{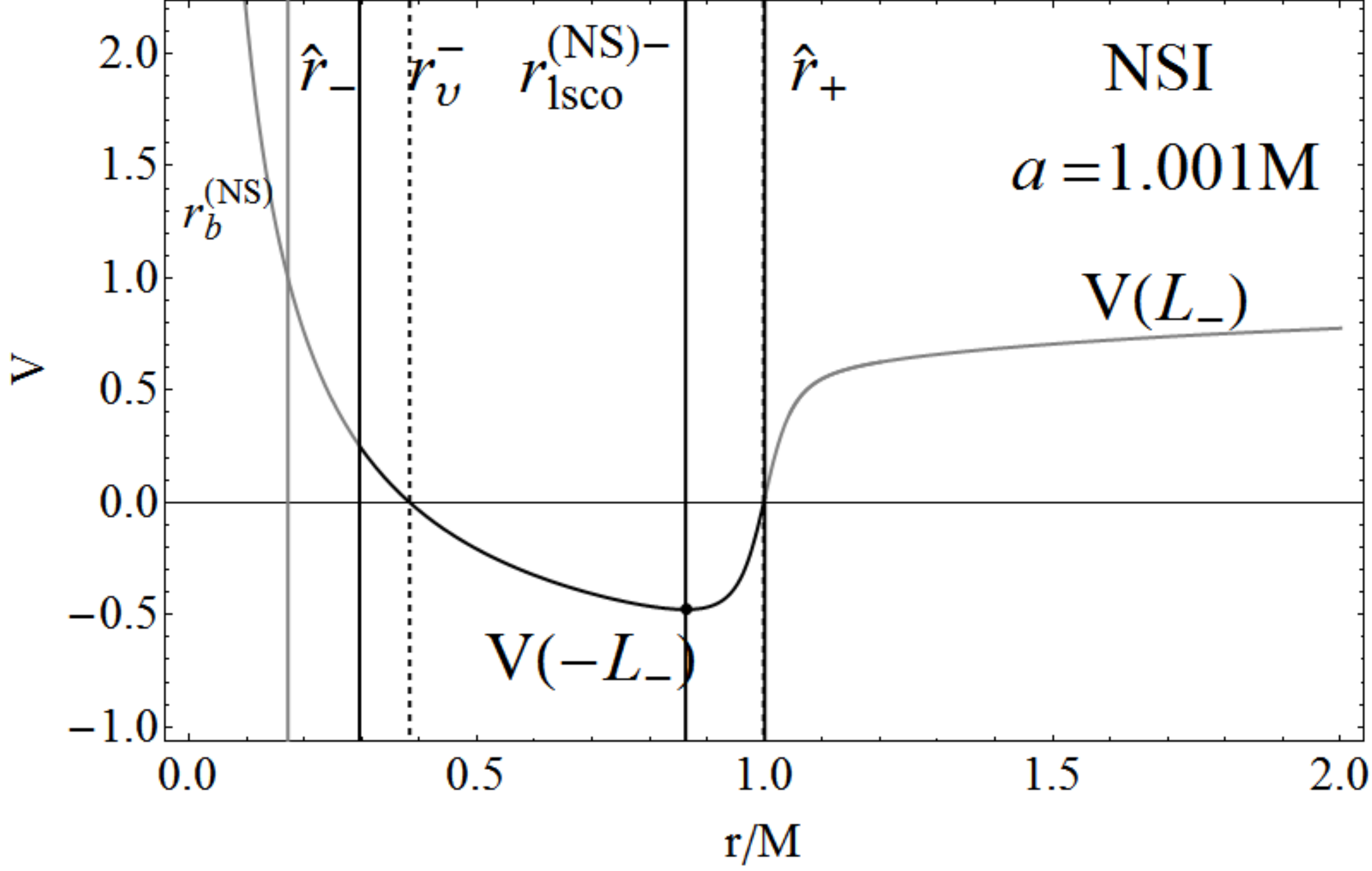}
\caption{\textbf{NSI} sources with $a\in]M,a_3]$. The energy $E_-^-\equiv V(-L_-)$  (black curves)  and $E_-\equiv V(L_-)$  (gray curves), in units of the particle mass $\mu$,  of the circular orbits for different spacetime spins. The last stable circular orbit $r^{(NS)-}_{lsco}$ il also plotted and the minimum of the energy is marked with  a point. At $a=a_3$ it is $\hat{r}_{\pm}=r^{(NS)-}_{lsco}$. The potential $V(-L_-)$  is plotted for the entire range $r\in]0,r_{\epsilon}^+]$, but the only counterrotating orbit is located at the boundary point $\hat{r}_{\pm}=r^{(NS)-}_{lsco}$. For spacetimes with $a\in]M, a_{\mu}[$ the energy of the circular  orbit is negative, and for a source with $a=a_{\mu}\equiv(4 \sqrt{{2}/{3}})/{3} M$ the energy vanishes, $E_-^-(r_{lsco}^{(NS)-},a_{\mu})=0$. }
\label{EnergycirNSIzoomm}
\end{figure}

Naked singularities  of the \textbf{NSI} class present a  rather rich and articulated dynamical structure of the ergoregion, as shown  in Table\il\ref{Table:NS}.
As they limit with the \textbf{extreme-BH} case at $a=M$,   \textbf{NSI} spacetimes  could be involved  in the hypothetical transitions between the   \textbf{BH} and  \textbf{NS} classes.
In contrast with the \textbf{BH}-spacetimes,  these geometries are characterized by two types of circular orbits: the corotating orbits  with momentum  $L=L_-$, which extend to the  regions $r>r_{\epsilon}^+$, and the counterrotating ones with  $L=-L_-$, confined in  a bounded orbital  region $\Sigma(-L_-)\subset\Sigma_{\epsilon}^+$  with boundaries at $r=\hat{r}_{\pm}$, as defined in Eq.\il(\ref{rpm}),  corresponding to  circular orbits with zero angular momentum $L=0$.
The orbital region $\Sigma(-L_-)$ is split by the radius  ${r}^{(NS)-}_{lsco}$ so that  $\Sigma(-L_-)=\Sigma_u^<(-L_-)\cup\Sigma_s(-L_-)$.
 These regions do not intersect each other, i.e., $\Sigma(-L_-)\cap\Sigma(L_-)=\emptyset$,  in contrast with the region $r>r_{\epsilon}^+$, where it is  possible to have, in the same orbital region, corotating and counterrotating matter. The regions $\Sigma(-L_-)$ and $ \Sigma(L_-)$ are  completely disjointed in
$\Sigma_{\epsilon}^+$,  although the presence of circular orbits with $L= 0$ on $\hat{r}_{\pm}$ could suggest a continuous transition between the two regions.
     For orbits with $L=0$ on $r=\hat{r}_\pm$, the particle energy is positive and   increases with the  source spin, with the condition
 $E({\hat{r}_-})>E({\hat{r}_+})$ (see Figs.\il\ref{Exremenagr} and \ref{EnergycirNSIzoomm}).
The existence of counterrotating orbits can be seen as a  ``repulsive gravity''  effect  that has been detected also in other  naked singularity spacetimes
\cite{Pu:Charged,Pu:Neutral,Pu:KN}, creating therefore an  ``antigravity''
sphere bounded by orbits with zero angular momentum. The stability of these orbits would suggest the presence of a belt of counterrotating material covering the singularity.

In  $\Sigma(-L_-)$, there is a class of naked singularities $\mathbf{NSIa}\equiv a\in]M,a_{\mu}]$
 where circular counterrotating orbits are possible  with negative energy,   ${\Large{\not} \Sigma}={\Large{\not}\Sigma_{s}}(-L_-)
      \cup{\Large{\not}\Sigma_{u}^<}(-L_-)$.  The energy of the particles in  these orbits is always less than $\mu$ in magnitude.
			The region  ${\Large{\not} \Sigma}$  is bounded by the  orbits $r_{\upsilon}^\pm$, {zero energy orbits} (see Fig.\il\ref{rewpoettx}).  Counterrotating particles are confined inside the  closed region  $\Sigma_s(-L_-)\cup\Sigma_u^<(-L_-) $ in $\mathbf{NSIa}\cup \mathbf{NSIIb}$ with a section of  negative energy particles
    ${\Large{\not}\Sigma_{s}}(-L_-)
      \cup{\Large{\not}\Sigma_{u}^<}(-L_-)$ in $\mathbf{NSIa}$.
The  symbol  ${\Large{\not}} \Sigma$ is used  in relation with the   regions of negative energy orbits   and, unless otherwise specified, $\Sigma$ will be considered for positive energy orbits, while  the labels $(\geq)$ and $(<)$ are to be understood, in the case of negative energies, related to the energy magnitude.

The existence of stable and unstable circular orbits with  $L<0$ and $E<0$, although located on an orbital region far from the source, can be important for the phenomena of accretion from the equatorial plane, because it would imply dropping  ``test"  material into the singularity with a negative contribution to the total energy and momentum.
The   region of stable orbits   is disconnected  in the sense discussed previously in \cite{Pu:Kerr,Pu:Neutral,Pu:Charged,Pu:Class,Pu:KN}, i.e.,
 if we imagine a hypothetical accretion disk made of test particles only, the disconnected stability regions form a ring-like configuration around the central object
 \cite{Pu:Neutral,Pu:Charged,Pu:Class}.

The existence of the region  $\Large{\not{\Sigma}}$
  is an intrinsic characteristic of \textbf{NS}-sources   that  has been also highlighted for  other axisymmetric exact solutions of the  Einstein equations, in particular, for the electrovacuum  spacetime described by the Reissner-Nordstr\"om (RN) solution\footnote{{In general, even if the  {global} structure of the Kerr spacetime is $\theta$-dependent,  on the equatorial plane these two axisymmetric solutions have some remarkable similar  geometrical features. For instance, in  the expressions for the outer and inner horizons $r_{\pm}$, where the horizons of the one solution can be fit into the outer by replacing    the spin parameter $a/M$ of the Kerr spacetime with the    electric charge parameter $Q/M$ of the RN solution.
Finally,
on the equatorial plane  the conformal diagram for the maximally extended Kerr (\textbf{BH} and \textbf{NS}) spacetimes is  identical to that of the RN solution \cite{GriPod09}.}}.
In this spacetime, the radii where $E=0$ define    the   \emph{effective ergoregion}  of the RN solution \cite{RuRR} or  $]r_{+}^{\ti{RN}}, r^{\epsilon}_{eff}[$, where $r_{+}^{\ti{RN}}\equiv M+\sqrt{M^2-Q^2} $ is the outer horizon in the RN geometry and {\small$r^{\epsilon}_{eff}\equiv M+\sqrt{M^2-Q^2(1-q^2/\mu^2)}$}, where $Q$ is the source charge and $q$ is the test particle charge. The existence of such a region  is  due to the  attractive electromagnetic  interaction    between the two charges
$(Qq<0)$,  resulting  in  negative energy states for test particles \cite{Pu:Charged,Pu:Class}. However, since the RN solution is  static, there is no ergoregion in this spacetime.

The  definition  of an effective ergoregion is introduced  in the description of the energy extraction phenomena of black holes, and can be extended also to the case of naked singularities. Therefore, one can define also for the \textbf{NSIa}  singularities the radii $r_{\upsilon}^{\pm}(a)$,  where $E=0$,
as the  \emph{outer effective ergosurface}  $r_{\upsilon}^{+}$ and  an \emph{inner effective ergosurface} $r_{\upsilon}^{-}$. Then, the region $]r_{\upsilon}^{-},r_{\upsilon}^{+}[$ would correspond to an \emph{effective ergoregion},
 where the   spin-orbit coupling terms in the Kerr source  plays the role of  the electrodynamic interaction between the test charge $q$ and the intrinsic charge of the  source in the definition of the effective ergoregion {in the RN geometry}.
However, {in the Kerr spacetimes,} the radii $r_{\upsilon}^{\pm}(a)$ are independent of the orbital angular momentum.

\subsubsection{The class $ \textbf{NSII}:
 a\in]{a_3},{a_4}]$}
The dynamical structure of $\Sigma_{\epsilon}^+$ in the   \textbf{NSII} geometries is shown in Table\il\ref{Table:NS} and in  Figs.\il\ref{rewpoett} and
\ref{rewpoettx}.  There are only corotating orbits with  $L=L_->0$ that can be  stable or unstable, bounded and  unbounded, so that the structure of $\Sigma_{\epsilon}^+$ is similar to the case of \textbf{BHIII} as given in Table\il\ref{Table:BHSIGMA1}. In the $\mathbf{NSII}$ class, the region $\Sigma_u^{\geq}(L_-)$ extends up to the singularity and there is no region  $\Sigma_{\varnothing}$. A further difference with   \textbf{BHIII} spacetimes is that the length $\mu_s$ of   $\Sigma_{s}$ in \textbf{NSII} decreases with the  attractor spin. This can be interpreted by saying that while higher spins in the  \textbf{BH} geometries favor  the orbital stability, an increase of the attractor spin in the \textbf{NS} spacetimes acts in the opposite direction,  favoring the instability of the (unbounded) orbits.
A pressure supported accretion disk  can be then entirely contained in $\Sigma_{\epsilon}^+$, and eventually evolve towards the instability,  moving the inner edge of the disk towards the region $\Sigma_{u}^{<}$, and eventually towards  $\Sigma_u^{\geq}$, giving raise to funnels of materials  launched outwards from the minimum point of the hydrostatic pressure. The length of the region where the center of the closed configuration can be located does not exceed $\mu_{s}=1.25M$,  while it decreases to zero as the singularity spin reaches  the upper limit  $a_4$ of this class of sources. Moreover, in the wider region, where a fixed attractor is in $\Sigma_u^<$ with a minimum length of $\mu_u^<\approx 0.48M$ and maximum  of $\mu_u^<\approx 1.085M$, there are toroidal  surfaces close around the maximum of the  pressure and  points of instability in $\Sigma_u^<$ from which there could be an overflow of matter.
As shown in Figs.\il\ref{Exremenagr}, the orbit energies  and the angular momenta increase with the spin of the source,  and can reach a minimum on the static limit
(see Sec.\il\ref{sec:stl}).
In order to set  test particles in circular orbits in these spacetimes we should provide a specific energy that increases with the attractor spin.
\subsubsection{The class $ \mathbf{NSIII}: a\in]{a_4},+\infty[$}
 In  \textbf{NSIII} geometries, as in the \textbf{BHII} class of attractors, the  marginally bounded orbit $r_b^{(NS)}$ crosses  the static limit as the spacetime spin is  $a=a_b^{NS}   \equiv 2 \left(1+\sqrt{2}\right)M\in\textbf{NSIII} $ (see Table\il\ref{Table:NS} and Fig.\il\ref{rewpoettx}).
It is therefore convenient to introduce a further decomposition of this class into two sub-classes,   namely $\mathbf{NSIIIa}$ where $a\in]a_4,a_b^{NS}]$ and $\mathbf{NSIIIb}$ where $a>a_b^{NS}$.
In all $\mathbf{NSIII}$ spacetimes,
the stable orbits are  suppressed for    $\mathbf{NSIIIa}$  attractors,   but    bounded unstable orbits are still possible.
In the   $\mathbf{NSIIIb}$  spacetimes  the ergoregion can be filled only  by  unbounded orbits
up to the  singularity, and the dynamical structure  of the ergoregion reduces to $\Sigma_{\epsilon}^+=\Sigma_{u}^{\geq}$.
In the  $\Sigma_u $ regions,   the   particle orbits are unstable and, correspondingly, extended pressure supported  toroidal matter  orbiting around these attractors have   instability  points  located  in $\Sigma_{u}^<$, where an overflow of matter is possible from a closed configuration. Moreover, there can also be Jet launches  with  funnels of materials in  $\Sigma_u^{\geq}$, which characterizes in particular  all the ergoregion of the $\mathbf{NSIIIb}$ spacetimes.

On the other hand, only in the  $\mathbf{NSIIIa}$ geometries the disk center can be located  in the region $r>r_{\epsilon}^+$ close to the static limit.
In this respect, the dynamical structure of  $\Sigma_{\epsilon}^+$ in   $\mathbf{NSIIIa}$ and $\mathbf{NSIIIb}$ spacetimes  is similar to the $\mathbf{BHIIb}$  and  $\mathbf{BHIIa}$ classes, respectively.

 We can conclude that naked singularities with  sufficiently high  spins,   \emph{strong naked singularities} of the $\mathbf{NSIIIa}$  class, and black holes
with sufficiently small spin within the  $ \mathbf{BHIIb}$ class do   not allow any  stable orbiting  matter  in the ergoregion. The orbital energy and angular momentum  of the circular orbits in $\mathbf{NSIII}$ spacetimes increase with  the singularity spin, as shown in  Figs.\il\ref{Exremenagr}. Bounded orbits are not allowed in \emph{very strong naked singularities} of the  $\mathbf{NSIIIb}$ class, but they can exist in \emph{weak black holes}  of the  $\mathbf{BHIIa}$ class with lower spin (see Fig.\il\ref{rewpoett}). An increase of the attractor spin in  weak black hole geometries, or a decrease in strong and   very strong naked singularities, affects {positively} the process of formation of a  circular orbiting configuration.
\section{Conclusions}
\label{sec:con}
In this work, we performed a complete analysis of the properties of circular motion inside the ergoregion of a Kerr spacetime. From the physical point of view, the  {ergosurface} is a particularly interesting surface, because it represents the limit at which an observer can stay at rest. In the ergoregion, an observer is forced to move due to the rotation of the gravitational source.

The dynamics inside the ergoregion is relevant    in Astrophysics for the  possible observational effects,  as the matter can eventually be captured by the  accretion,  increasing or removing part of its energy and angular momentum,  prompting a shift of its spin, and inducing  an  unstable phase  where the intrinsic spin  changes
(spin-down and spin-up  processes with the consequent change in the causal  structure). {A discussion on the ergoregion stability can be found in \cite{Cardoso:2007az,CoSch}. For further consideration of a possible destruction of the horizon and naked singularity formation see, for example, \cite{Li:2013sea,Jacobson:2010iu,Stu80,BiStuBa89,BaBiStu89,Kovar:2010ty}.}
 It is therefore possible that, during the   evolutionary phases of  the rotating object, the interaction with orbiting matter could lead to evolutionary stages of spin adjustment, in particular  for example  in the proximity of the extreme Kerr solution (with $a\lesssim M$) where  the speculated spin-down of the \textbf{BH} can occur preventing the formation of  a naked singularity with $a \gtrsim M$
(see also \cite{Esitenza,
Gao:2012ca,
Evo,
Stuchlik:2011zza,
vanPutten,
Gammie:2003qi,
Abo,
Kesden:2011ma,
Wald74,
J-S09} and \cite{Pradhan:2012yx}). On the other hand, the accreting matter  can even get out, giving rise, for example, to Jets of matter or radiation \cite{Meier}.
{However, it is important to emphasize that different processes can lead to Jets emission from black hole accretion disks;
 in general, it is widely believed that a crucial role is played by the electromagnetic field,  giving rise to  magnetically driven Jets. Another possibility is the extraction of  energy from a rotating black hole  through the Blandford-Znajek mechanism. See, for example, \cite{RZN,PVP,P11,AO02,PCGG,BSZ10,FS10,IHK85,SKC,HHJEPJC} for general  discussions on the   role of magnetic fields in the formation of Jets.
 A possible generalization of the present work to include this mechanism would imply a  thorough  study of   the effects of the
 electromagnetic field on the particle acceleration
in the spacetime of a rotating black hole.
Several aspects  of this problem have been addressed in the literature, for example,  in \cite{BZfirst,TPM86,SNT15,ZTN14}  where it  was  investigated  how the
 magnetic field affects  the motion of test particles in the vicinity of a black hole.
Further discussions on the role of the  electromagnetic fields in  black-hole physical processes  can be found  for example in \cite{Pen-Dan183,MPDad}}.

In our analysis, we limit ourselves to the study of circular orbits located on the equatorial plane of the Kerr spacetime.
On the equatorial plane, the static limit  does not depend on the source spin,  but for any Kerr spacetime it is $r_{\epsilon}^+=2M$ and it  coincides with the event horizon of the Schwarzschild spacetime. This is a simple setting that  allows an immediate comparison with the limiting case $a=0$.
 Furthermore, matter configurations in accretion are typically  axially symmetric and many of the  geometrical and dynamical characteristics of the disk are determined by the properties of the configuration on the accretion equatorial section.

Our approach consists in rewriting the geodesic equations in such a form that the motion along circular orbits is governed by one single ordinary differential equation whose properties are completely determined by an effective potential. The conditions imposed on the effective potential for the existence of circular motion allow us
to derive explicit expressions for the energy and angular momentum of the test particle. The behavior of these physical quantities determine the main properties of the circular orbits in terms of the radial distance which, in this case, coincides with the radius of the orbit, and the intrinsic angular momentum of the gravitational source.
We performed a very detailed investigation of all the spatial regions inside the ergoregion where circular motion is allowed. In addition, we investigate the stability properties of all the existing circular orbits.

The distribution of circular orbits inside the ergoregion turns out to depend very strongly on the rotation parameter $a$ of the source, and makes it necessary to split the analysis into two parts: black holes and naked singularities. In addition, the behavior of the effective potential in the ergoregion in terms of the rotational parameter suggests an additional split by means of which black holes become classified in four classes, namely, \textbf{BHI}, \textbf{BHIIa},  \textbf{BHIIb}, and
\textbf{BHIII}, whereas naked singularities are split into five classes, namely, \textbf{NSIa},  \textbf{NSIb},  \textbf{NSII},  \textbf{NSIIIa}, and  \textbf{NSIIIb}. We then investigate in detail for each class the behavior of the energy and angular momentum of the test particle, as well as the properties of the effective potential. In this manner, the analysis of circular motion allows us to derive physical information about entire sets of black holes and naked singularities.

Circular motion is possible inside the ergoregion in black holes and naked singularities as well. However, there are fundamental differences if we consider the stability properties. In the case of black holes, only the set \textbf{BHIII} can support a spatial region with particles moving along circular corotating stable orbits. The \textbf{BHIII} class includes all the black holes whose rotation parameter is contained within the interval $a/M \in ]2\sqrt{2}/3,1]$, i.e., rapidly rotating black holes  including the extreme black hole. The spatial region with stable particles extends from the radius of the ergoregion ($r_\epsilon^+ = 2M)$ to the radius of the last stable circular orbit, so that the maximum radial extension of this region is $M$ for an extreme black hole, and the minimum extension is zero for a black hole with $a= 2\sqrt{2}/3M$. The last case corresponds to particles moving on the last stable circular orbit. If we imagine a hypothetical accretion disk made of test particles only, then we conclude that black holes can support inside the ergoregion only one corotating disk with a maximum extension of $M$.
Introducing the concept of dynamical structure of the ergoregion, we discussed also the properties of  toroidal extended configurations  (pressure supported hydrodynamic accretion disk models)  in the  regions that determine the dynamical structure. This structure is formed by a (uniquely defined) collection of disjoint regions of the orbit plane  whose union recomposes the ergoregion. A change in the orbit, at fixed spin, can lead to a shift in the dynamical structure, while a change of  the attractor spin can lead to a transition between the regions; moreover, a shift in the  boundary spins  is equivalent to a transition between classes and sometimes  the regions of the structure.
The stability  analysis,  with respect to   a change in the spin of the source, has led to the identification of special \textbf{BH} sources (with $a={a}_b^-/M\approx0.828427$ and $a= a_2/M \approx0.942809$)   where a possible change in the value of the spin, in the margins between the different regions of the dynamical structure, affects appreciably  the properties of the  toroidal structures, {indicating that this kind of structures are
more sensitive to a change of the source.}

The case of naked singularities is more complex. In fact, one of the  interesting features is that inside the ergoregion there can exist particles moving along counterrotating stable orbits. As a consequence, the location and structure of the regions with stable particles is much more complex than in the case of black holes. In particular, it was shown that there can exist several stability regions, separated by instability regions. This implies a discontinuous structure of the stability regions so that, if certain energetic conditions are satisfied,  an accretion disk
made of test particles would show a ring-like structure. This makes naked singularities essentially different from black holes. The
characteristics  of the rings and their extensions depend on the explicit value of the rotation parameter.  Finally, we found that there exists a maximum value of the spin for which no more stable configurations can exist, namely, $a=2\sqrt{2}M \approx 2.828M$  where the radius of the last stable circular orbit coincides with the radius of the ergoregion. Naked singularities with spins greater than this critical value do not support any  {stable} disk-like or ring-like configurations {entirely contained in  }  the ergoregion.
In \textbf{NSI} spacetimes,  there are both stable and unstable counterrotating   orbits inside the ergoregion.
This fact can be understood as the effect of repulsive gravity, but it is interesting to note that this phenomenon can occur  only in  sufficiently slow rotating naked singularities with spin values close to the value of the \textbf{extreme BH}-case.
 In \textbf{NSII} spacetimes,  there can exist stable corotating orbits; this is the major difference with the \textbf{BH} case   and  it represents also the main difference with the other  \textbf{NS} sources.
We have  introduced the concept of
\emph{inner and outer effective ergosurface}   defined by the radii $r_{\upsilon}^{-}$ and $r_{\upsilon}^{+}$, where $E=0$.
The effective ergoregion (at $r\in]r_{\upsilon}^{-},r_{\upsilon}^{+}[$) is defined for supercritical configurations with $a\in]M, 1.08866M]$ in \textbf{NSI} spacetimes,  where  $E<0$.
Our results show that the complex stability properties of circular orbits inside the ergoregion of naked singularities is due to the presence of effects that can be interpreted as generated by repulsive gravitational fields. The nature of this type of fields is not known. We expect to investigate this problem in future works by using certain invariant properties of repulsive gravity \cite{lq14}.
For the sake of completeness, we also investigated  all the properties of circular orbits in the limiting case of extreme black holes, and classify all the sources that allow circular orbits on the radius of the static limit. In both cases, we used the available physical parameters to perform a detailed analysis  confirming  the rich structure of the gravitational sources described by the Kerr spacetime.

\begin{acknowledgements}
This work was supported in part by DGAPA-UNAM, grant No. 113514, and Conacyt, grant No. 166391. DP gratefully acknowledges financial support from  Blanceflor Boncompagni-Ludovisi, n\'ee Bildt,  and would like to thank the institutional support
of   the Faculty of Philosophy and Science of the Silesian University of Opava. DP thanks an anonymous colleague who has read in great detail the work  providing important suggestions  to improve the  presentation of the results.
\end{acknowledgements}

\appendix

\section{Stability of circular orbits and notable radii}
\label{Sec:po}
The \emph{last circular orbits} $r_{\gamma}\in\{r_{\gamma}^-,r_{\gamma}^+\}$, are located at
\bea\label{Eq:gamma}
{r_{\gamma}^{\mp}}\equiv2 M\left(1+\cos \left[\frac{2 \arccos\left(\mp \frac{a}{M}\right)}{3}\right]\right),
\eea
where $r_{\gamma}^-\in \Sigma_{\epsilon}^+$ in the $\mathbf{BH}$ geometries.
The stability properties of the particle circular dynamics  are regulated  by the  second radial derivative of the effective potential: the saddle  points  of the function $V$, given  by the conditions $V'=0$ and
$
V'' =0 \ ,
$
 define the radii of the \emph{last stable circular orbits}  $r_{lsco}\in\{r_{lsco}^{\mp},r_{lsco}^{(NS)},r_{lsco}^{(NS)-}\}$ where
\bea\label{Eq:rlscomp}
r_{lsco}^{\mp}&\equiv& M\left(3+Z_2\mp\sqrt{(3-Z_1)(3+Z_1+2Z_2)}\right),\\\nonumber
 {r_{lsco}^{-}}&& {\mbox{for}\quad \mathbf{BH}: L=L_{-}},\quad  {r_{lsco}^{+}\quad \mbox{for}\quad \mathbf{BH}/\mathbf{NS}: L= -L_{+}}.
\eea
Here $r_{lsco}^{-}$ is for corotating with $L=L_-$ orbits in a black hole geometry and $r_{lsco}^{+}$ is for counterrotating orbits with $L=-L_+$ in black hole and naked singularity spacetimes,  respectively, and
\bea\nonumber
&&r^{\ti{(NS)-}}_{lsco}\equiv M\left[3-Z_2+\sqrt{(3-Z_1)(3+Z_1-2Z_2)}\right]
\\&&\label{Eq:rlscompNS}
\mbox{for}\quad a/M>1.28112 \quad\mbox{and}\quad L=L_{-},
\\\nonumber
&&
r^{\ti{(NS)-}}_{lsco}\equiv M\left[3-Z_2-\sqrt{(3-Z_1)(3+Z_1-2Z_2)}\right]\\
&&\mbox{for}\quad a/M\in]1, 1.28112[\quad\mbox{and}\quad  {L=- L_{-}}
\eea
where $a=1.28112M\in\mathbf{NSI}$\footnote{The spin $a=1.28112M$ in  Eq.\il(\ref{Eq:rlscompNS}) is introduced only for  convenience in  the definition  of the functions    $r^{\ti{(NS)-}}_{lsco}$ that  can be  matched for continuity in $a=1.28112M$.}  and $Z_2\equiv\sqrt{3(a/M)^2+Z_1^2}$ where $$Z_1\equiv1+\left[1-(a/M)^2\right]^{1/3}\left[(1+a/M)^{1/3}+(1-a/M)^{1/3}\right].$$ It is $(r_{lsco}^-,r_{lsco}^{(NS)-})\in \Sigma_{\epsilon}^+$.

The \emph{marginally bounded orbits},  $r_b\in\{r_b^{\pm}, r_b^{(NS)-}\}$, are located at
\bea\label{Eq:perd-BH}
&&r_b^{\pm}\equiv2M\pm a+2 \sqrt{M}\sqrt{M\pm a},\\\nonumber&&
{r_{b}^{-}\quad\mbox{for}\quad \mathbf{BH}:L= L_{-}},\quad {r_{b}^{+}\quad \mbox{for}\quad \mathbf{BH}/\mathbf{NS}: L=-L_{+}}
\eea
for particles  corotating with angular momentum $L=L_-$  around a  black hole attractor, and  for  counterrotating particles  with momentum  $L=-L_+$ in  black hole and naked singularity geometries respectively, and
\be\label{Eq:perd-NS}
r_b^{(NS)-}\equiv 2M + a - 2\sqrt{M} \sqrt{M + a}\quad\mbox{for}\quad \mathbf{NS}:\; L=L_-
\ee
 for particles with   $L=-L_-$  orbiting in a naked singularity geometry. It is $\{r_{b}^-,r_b^{(NS)-}\}\in\Sigma_{\epsilon}^+$.

The radii $\hat{r}_{\pm}\in \Sigma_{\epsilon}^+$,  introduced in Sec.\il(\ref{sec:nss}),  for the  \emph{circular orbits with zero angular momentum} are solutions of the equation  ${V}'=0$ with $ L=0$:
\bea\nonumber
\hat{r}_{\pm}&\equiv&\frac{1}{\sqrt{6}}\left[\mathfrak{S} \pm\sqrt{\frac{6\sqrt{6}
a^2M}{\mathfrak{S}}-\mathfrak{S}^2-6a^2}\right],\quad
\mathfrak{S}\equiv\sqrt{\frac{4a^4}{\mathfrak{s}^{1/3}}+\mathfrak{s}^{1/3}-2a^2},
\\\label{rpm}
\mathfrak{s}&\equiv&\left(27M^2a^4-8a^6+3M\sqrt{81M^2a^8-48a^{10}}\right)\ ,
\eea
in the   naked singularity geometries   of $\mathbf{NSI}$-class (see also \cite{Pu:Kerr}). In the case of \textbf{NSIa} spacetimes, there are also orbits $r_{\upsilon}^\pm\in\Sigma_{\epsilon}^+$ of counterrotating particles with  $L=-L_-$  and zero energy $\mathcal{E}=0$:
\bea\nonumber
r_{\upsilon}^+&\equiv&\frac{4}{3} M\left(1+{\sin}\left[\frac{1}{3} \arcsin\left(1-\frac{27 a^2}{16M^2}\right)\right]\right),
\\\label{Eqs:rnu0}
r_{\upsilon}^-&\equiv&\frac{8}{3} M\sin\left[\frac{1}{6}\arccos\left(1-\frac{27 a^2}{16M^2}\right)\right]^2.
\eea
\section{Spacetimes classes and limiting spins}
\label{sec:ebh}
\begin{figure}%
\begin{tabular}{cc}
\includegraphics[scale=.3]{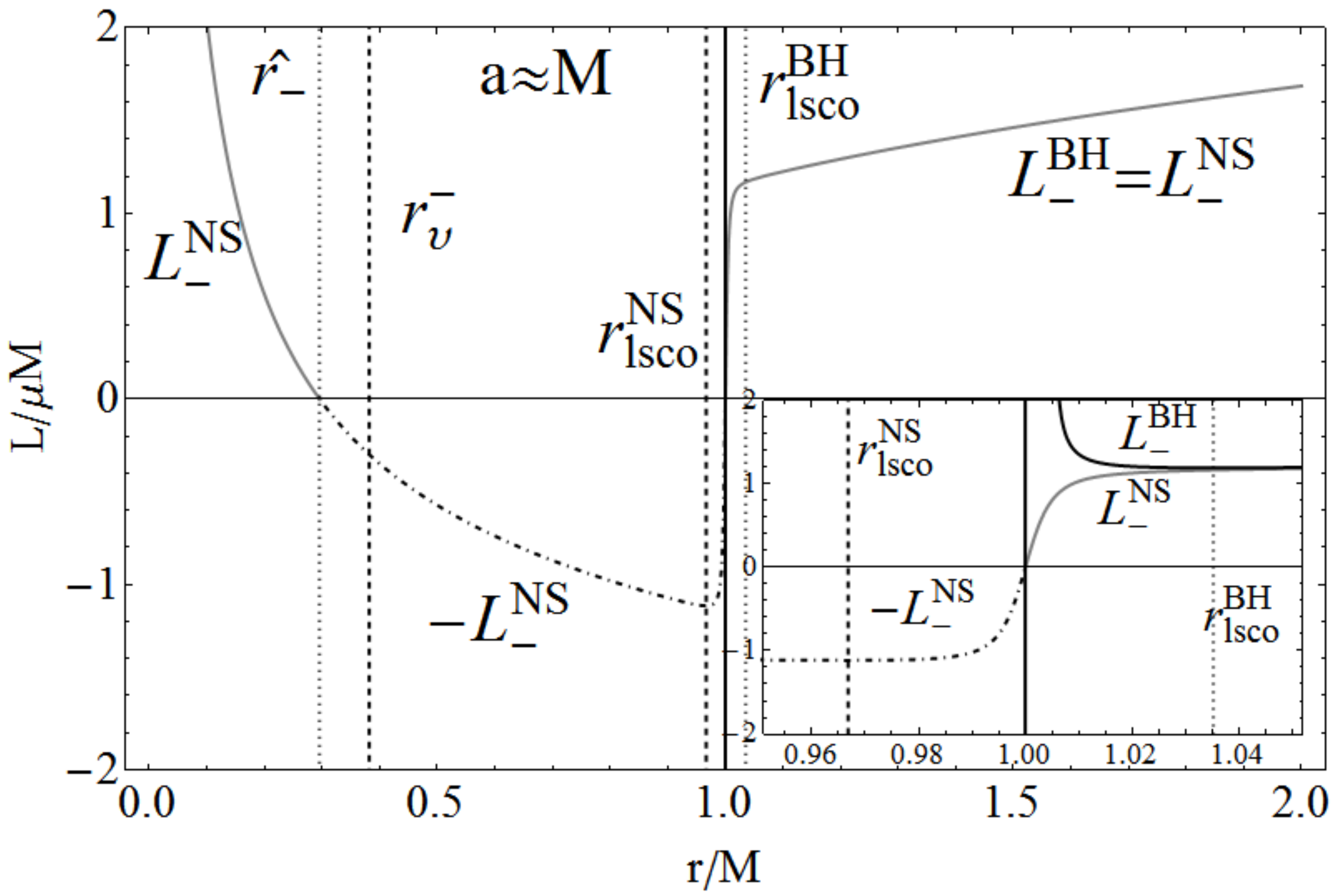}
\\
\includegraphics[scale=.3]{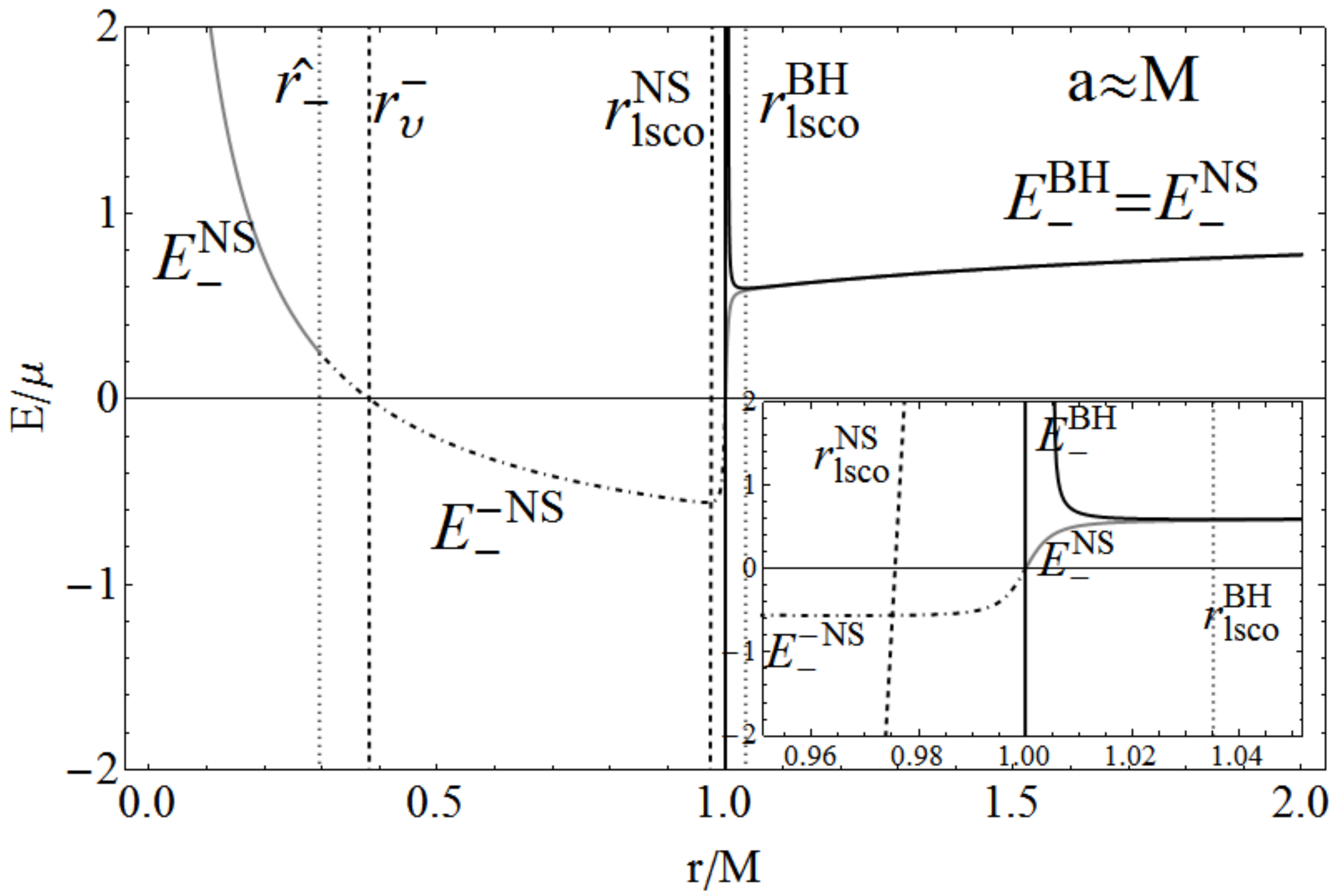}
\end{tabular}
\caption[font={footnotesize,it}]{\footnotesize{ Orbital energies and angular momenta in the \textbf{BH}-case,  with $a/M=1-\epsilon$, and the \textbf{NS}-case, with $a/M=1+\epsilon$ and $\epsilon=10^{-5}$. To simplify the readout of the plots, we adopted a notation distinguishing  explicitly the  \textbf{NS} and \textbf{BH} cases:   $E_-^{-NS}\equiv E(-L_-)$  and $E_-^{NS}\equiv E(L_-)$, for the naked singularity, $E_-^{BH}$ and $L_-^{BH}$ for the black hole. The marginally stable orbits are denoted  as $(r_{lsco}^{BH},r_{lsco}^{NS})$. The orbit $\hat{r}_-$, where $L=0$, is shown with a dashed line. It is $r_{lsco}^{{BH}}=1.03523M$, $r_{lsco}^{NS}=0.966815M$, and $\hat{r}_{+}=M$, $\hat{r}_-=0.295605M$, {inside plots are zooms for the region close to the \textbf{BH} horizon}.}}
\label{Exremenagr1}
\end{figure}

\begin{figure}
\begin{tabular}{cc}
\includegraphics[scale=.3]{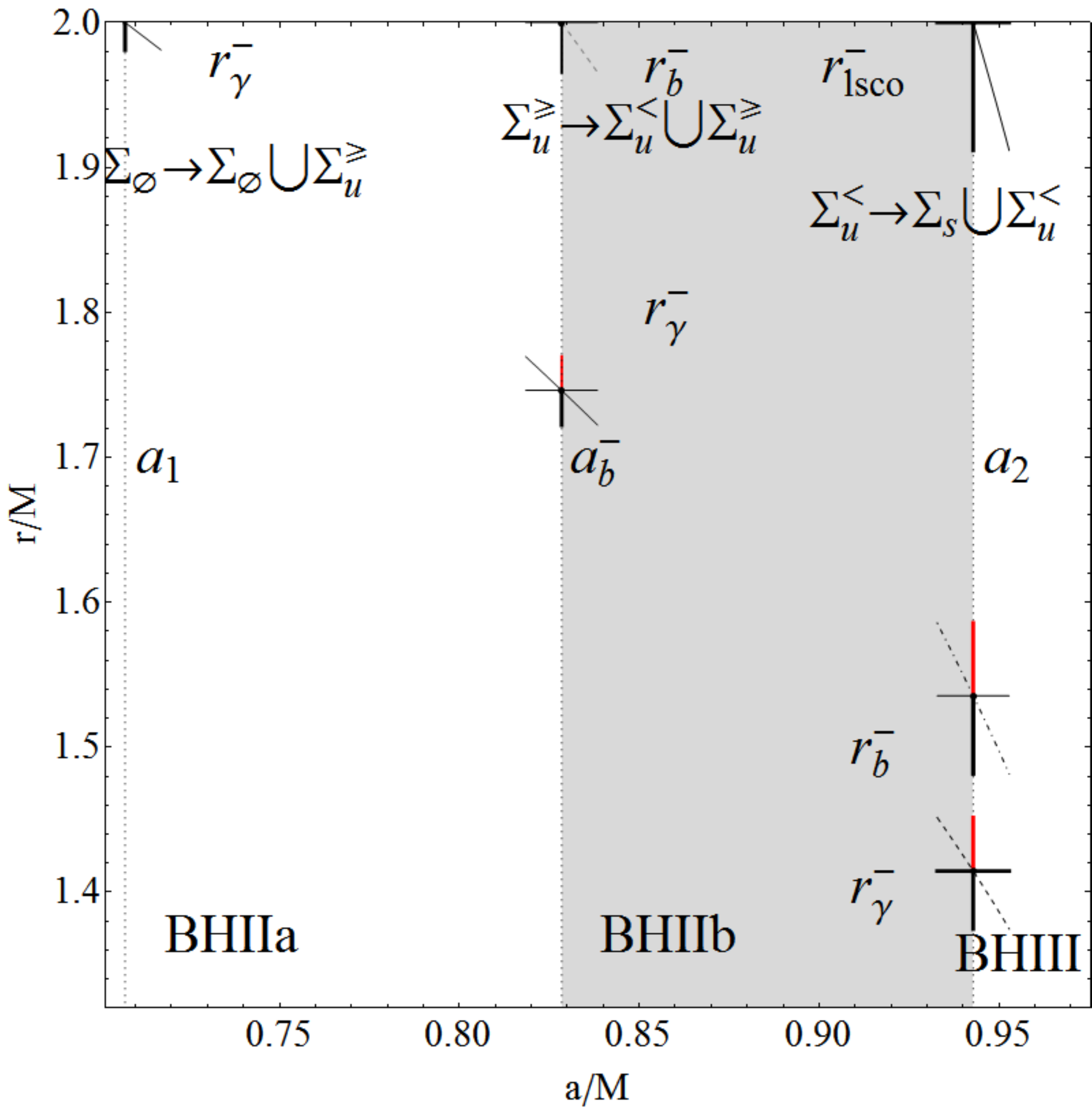}
\end{tabular}
\caption[font={footnotesize,it}]{\footnotesize{Black hole geometries in the  region $ \Sigma_{\epsilon}^+ $. The \textbf{BH} spin $a$  and the radius $r$ are in units of mass $M$. The \textbf{BHIIa}, \textbf{BHIIb} and  \textbf{BHIII} classes are explicitly shown. Dotted lines represent the  spins $a_i\in \{a_1,a_b^-,a_3\}$. The points correspond to
the couples  $(r_{\gamma}^-(a_1),a_1)$, where $r_{\gamma}^-(a_1)=r_{\epsilon}^+$,
$(r_{\gamma}^-(a_b^-),a_b^-)$ and  $(r_b^-(a_b^-),a_b^-)$, where $r_b^-(a_b^-)=r_{\epsilon}^+$,
$(r_{\gamma}^-(a_2),a_2)$,  $(r_b^-(a_2),a_2)$, and
$(r_{lsco}^-(a_2),a_2)$,  where $r_{lsco}^-(a_2)=r_{\epsilon}^+$.
The horizontal black lines  crossing the points are $a_i\pm\epsilon_a$,
where $\epsilon_a=1/100M$. The vertical black lines are $r_i(a_j+\epsilon_a)$ and the red vertical lines are $r_i(a_j-\epsilon_a)$.
The curves $r_i\in\{r_{\gamma}^-, r_{b}^-, r_{lsco}^-\}$ in the  ranges $a_i\pm\epsilon_a$ are plotted.
Close to the limiting couples
$(r_{\gamma}^-(a_1), a_1)$,  $(r_b^-(a_b^-),a_b^-)$
 and
$(r_{lsco}^-(a_2),a_2)$, we show explicitly the  transitions between different sections of $\Sigma_{\epsilon}^+$.
The  relevant transitions are for  a change at $a_b^-=0.828427M$ and  $a_2=0.942809M$.  }}
\label{Fig:tostay}
\end{figure}

 Interacting  with the surrounding matter and   fields, an  attractor  can pass, during its evolution,   through stages of adjustment of its spin, {spin-down} or {spin-up} processes see also \cite{Esitenza,
Gao:2012ca,
Evo,
Stuchlik:2011zza,
vanPutten,
Gammie:2003qi,
Abo,
Kesden:2011ma,
Wald74,
J-S09}.
These phenomena will involve also  the  interaction of the accretor   with matter and fields  in the ergoregion (see, for example, \cite{Meier}).
Questioning about  the possible disruption  or formation of a horizon and  the  consequent formation and existence of a \textbf{NS}-spacetime, it is obviously important to consider the possibility of a transition,   through  the  \textbf{BHIII} and \textbf{NSIa} classes that could lead to a disruption of the event horizon. In this case, the matter dynamics in the \textbf{BHIII} and \textbf{NSIa} geometries with spins $a\approx M$ is especially relevant.
For $a=M$ the effective potential  is an increasing function of the radius orbits with   $L\leq0$
\cite{Li:2013sea,Jacobson:2010iu}.
There are circular orbits with $L=L_-\in]{2}/{\sqrt{3}},1.68707[\mu M$.
Figure\il\ref{Exremenagr1} shows the analogies and differences between the orbits in the \textbf{BH}-geometry with   $a/M=1-\epsilon$ and in the \textbf{NS}-geometry with $a/M=1+\epsilon$ and
$\epsilon=10^{-5}$. The radius
  $\hat{r}_+$ for naked singularities can be defined at any value $a=M+\epsilon$ with $\epsilon>0$; the black hole counterpart  $\hat{r}_+=M+f(\epsilon)$, where $f(\epsilon)<0$, can be easily evaluated and is of the order of   $(a-M)^2$. The  orbital structures  in the two cases $a\lessapprox M$ and $a\gtrapprox M$  are completely different. 	Moreover, the radius { $r_{\upsilon}^+$} has a maximum at $a=M$.
In Fig.\il\ref{Exremenagr1}, we show the  angular momenta and energies of the orbits in two different regions. Fig.\il\ref{Fig:tostay}  shows the effects of a transition of the black hole geometries among the classes defined  in  Table.\il\ref{Table:asterisco} for a   shift of the source spin in $a_i\in \mathcal{A}_{BH}$ of $\pm\epsilon=\pm0.01 M$.
The  more significant  change is in the structure of  $\Sigma_{\epsilon}^+$ determined  as a consequence of a  shift $\pm\epsilon_a$ of the spins $a_b^-=0.828427M$ and  $a_2=0.942809M$.
An analogous plot can be made for the case of a naked singularity; however, as can be easily seen in  Figs.\il\ref{rewpoettx}, the most relevant transitions between the sections of the dynamic structure of $\Sigma_{\epsilon}^+$ occur between  \textbf{NSIa}  and  \textbf{NSIb} classes for a  change in the geometry at $a=a_\mu$, with the resulting shift of the region   ${\Large{\not}} \Sigma(-L_-)$ into  $\Sigma(L_-)$
(for an increase of the attractor spin),  between the classes $\mathbf{NSIb}$ and $\mathbf{NSII}$ for a  change in the geometry at $a=a_4$,  and the transition from  $\Sigma(-L_-)$ to $\Sigma(L_-)$  and from $\Sigma_u^<(L_-)$ to $\Sigma_s(L_-)$.
A thorough study of these sources will be the subject of further work.

\section{The static limit}
\label{sec:stl}
{We consider here with more details  the role of the static limit on the determination of the dynamical structure of the ergoregion, especially in the  \textbf{NS} sector. In particular, we will analyze the behavior of the circular motion in the regions very close to  $r_{\epsilon}^+$,  varying  the spin of the attractor to determine its influence on the determination of the phases of equilibrium of the orbiting configurations. Particular attention is given to the analysis of the energy and momentum of the particle in orbit  in those regions, as they are relevant for  the orbital decay.
{We will show that  both  conserved quantities do not have a monotonous trend with spin, as the source evolves, implying  the decay of  the orbiting matter}.  This may favorite those geometries where the energy or the orbital momentum  are at  their  minimum.}

\begin{figure}
\begin{tabular}{cc}
\includegraphics[scale=.31]{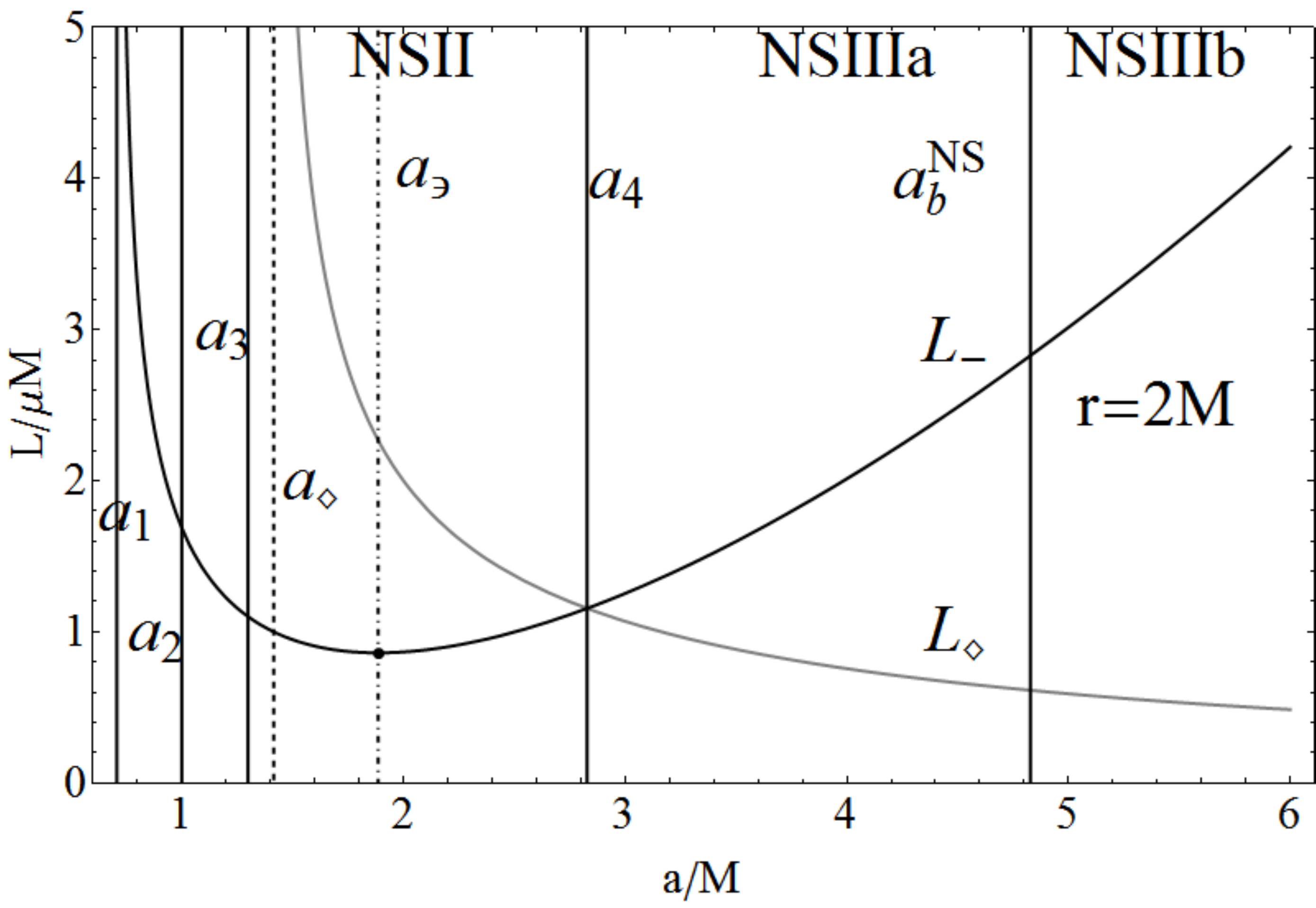}
\\
\includegraphics[scale=.31]{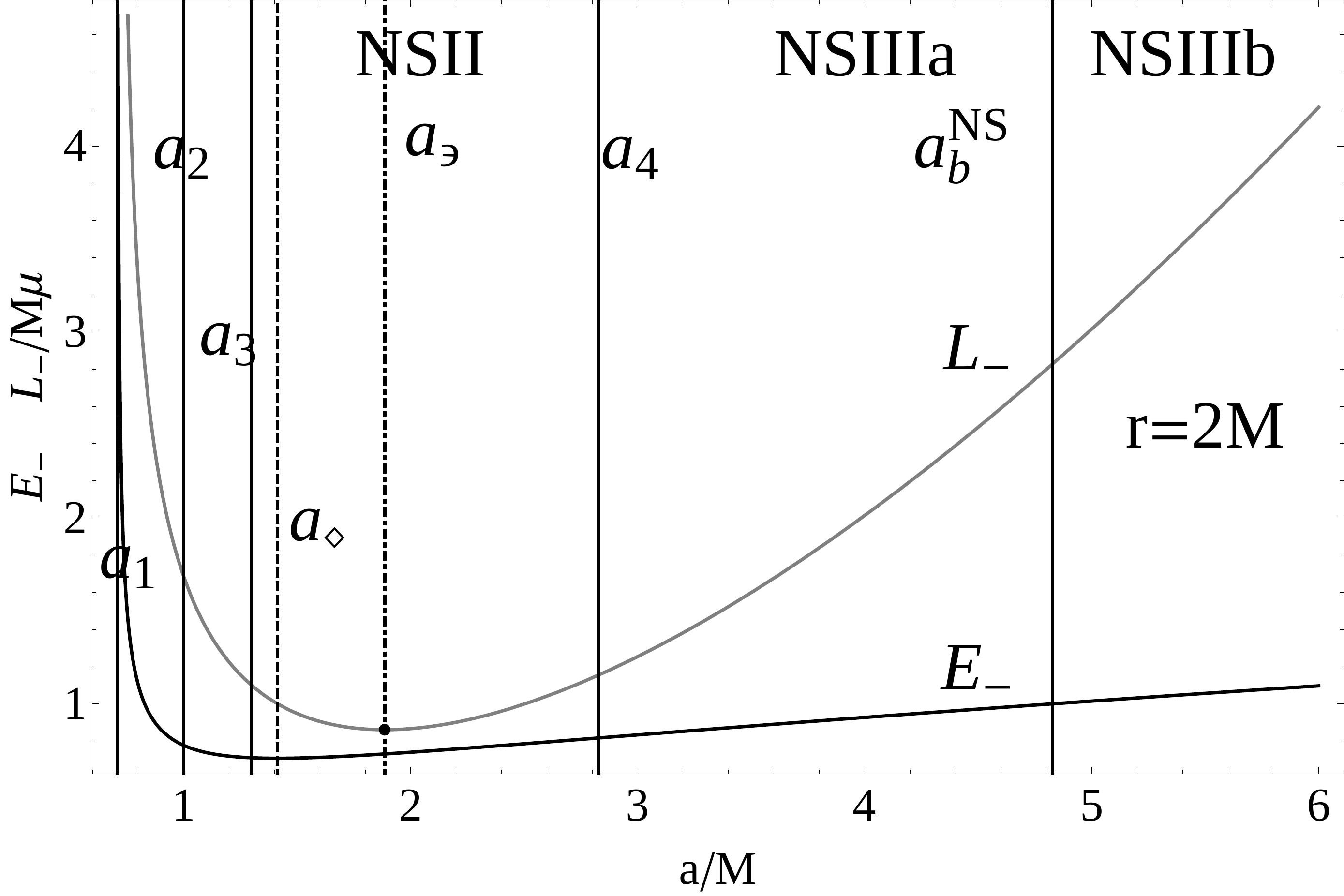}
\end{tabular}
\caption[font={footnotesize,it}]{Angular momentum and energy on the static limit $r_{\epsilon}^+=2M$. Bottom panel: the  orbital angular momentum $L_-$ (gray curve) and the energy $E_-$ (black curve) as functions of $a/M$. The minima are marked with points. The different \textbf{BH} and \textbf{NS} regions are also marked. It is $L_->E_-$. Upper panel: plot of $L_{\diamond}$ (gray curve) and $L_-$ (black curve) versus  the spin $a/M$. The asymptote at  $a_{\diamond}\equiv\sqrt{2}M$ is plotted with a dashed line. The spin $a_{\backepsilon}\equiv(4 \sqrt{2})/3 M\approx1.88562M$ is also plotted. At $L=L_{\diamond}$ the effective potential $V_{\epsilon}^+$ is constant with respect to a change of the attractor spin.}
\label{Fig:LogPlorru2LV}
\end{figure}

The effective potential function $V(r; L ,a )$ in given in Eq.(\ref{qaz})  is  well defined  and positive at $r=r_{\epsilon}^+$ for $a>0$,
 it has   an extreme  point on the static limit, according to  Eq.\il(\ref{Eq:Kerrorbit}),  for all  Kerr geometries, except for \textbf{BHI} spacetimes.
Therefore, at $r=r_{\epsilon}^+$ there is a circular orbit with angular momentum $L_{\epsilon}^+\equiv L_-(r_{\epsilon}^+)>0$ and energy $E_{\epsilon}^+\equiv V(r_{\epsilon}^+,L_{\epsilon}^+)>0$, but  in the spacetime with $a=a_1$ it is  $r_{\gamma}^-=r_{\epsilon}^+$,  that is, the static limit coincides with the photon orbit.
The timelike particle  orbit  $r_{\epsilon}^+$ is stable, i.e. $r_{\epsilon}^+\in\Sigma_s$, in \textbf{BHIII}, \textbf{NSI} and \textbf{NSII} geometries. Then, the static limit is a   {marginally stable circular orbit}  in spacetimes with spins  $a=a_2$ and $a=a_4$, while $r^+_{\epsilon}\in\Sigma_{u}^{\geq}$ in \textbf{BHIIa}  and  \textbf{NSIIIb}  spacetimes. Moreover, in the geometries in $a=a_b^-$  and $a=a_b^{NS}$ it is $r_{b}^-=r_{\epsilon}^+$ and  $r_{b}^{(NS)}=r_{\epsilon}^+$, respectively. On these orbits, the particle energy is $\mathcal{E}=E_-=\mu$, and the static limit coincides  with the marginally bounded orbit.
These special orbits should be interpreted as a limit of an orbit that approaches $ r_{\epsilon}^+$ from $\Sigma_{\epsilon}^+$ or the outer region.
An  observer at infinity  will verify that a particle  moving along an unstable  orbit will eventually cross the static limit and fall  into the source or escape away, depending on the attractor.
 The static limit on the equatorial plane is {independent} of the spacetime spin and therefore it can be considered as invariant with respect to a slow change of the  source spin (at constant $M$), but the constants of motion $ \mathcal{E} $ and $ \mathcal{L} $, instead, depend on the spin-mass ratio of the attractor and do not have a monotonically increasing behavior on the static limit with respect to a variation of the  source spin; instead, they each a minimum, as function of $a/M$, in  the \textbf{NSII} class of geometries.
The energy function in the static limit decreases with the source spin   for $a>a_{\diamond}$ and the particle angular momentum $L>L_{\diamond}$, where the
rotation parameter ${a}_{\diamond}\equiv\sqrt{2}M\approx1.41421M\in\mathbf{NSII}$, and the angular momentum
{
\be
\frac{L_{{{\diamond}}}}{\mu M}\equiv2 \sqrt{2} \sqrt{\frac{M^2}{a^2-2M^2}}\ ,
\ee
}
increases as the spacetime rotation decreases, and diverges at $a=a_{\diamond}$, as shown in Fig.\il\ref{Fig:LogPlorru2LV}.
When
$L=L_{\diamond}$, the potential  $V_{\epsilon}^+$ is constant, independently of the source spin.
For lower  spins,  $ a <a_{\diamond} $, regardless of the orbital angular momentum,  the energy function    increases with the spin of the \textbf{BH}- or
\textbf{NS}-sources.
In the spacetime  with  $a=a_{\diamond}$,  the particle energy on the static limit reaches a minimum value and, therefore, the particle  energy  increases as the spacetime rotation increases, until it reaches its asymptotic value for $a\rightarrow a_1$ (upper extreme of \textbf{BHI} sources), where  it is indeed $r_{\epsilon}^+=r_{\gamma}^-$ (see Fig.\il\ref{Fig:LogPlorru2LV}).
The particle orbital  angular momentum has a minimum, as a function of the spin-mass ratio, in   $a_{\backepsilon}\equiv2a_2\approx1.88562M\in \mathbf{NSII}$, where $a_3<a_{\diamond}<a_{\backepsilon}<a_4$.

%
%
%
%
%


\end{document}